\documentclass[elsart10,epsf,eqsecnum,graphics,cite]{revtex4}

\def\beq{\begin{equation}}
\def\eeq{\end{equation}}

\usepackage{amsmath,amsfonts,amssymb,amsthm,nccmath,latexsym,mathtools}
\usepackage[mathscr]{euscript}
\usepackage{epsfig,epsf}
\usepackage{graphicx}
\usepackage{grffile}
\usepackage{xcolor}
\usepackage{hyperref}
\usepackage{color} 

\begin{document}


\voffset1.5cm

\title{The Color Glass Condensate density matrix: \\ Lindblad evolution, entanglement entropy and Wigner functional}

\author{N\'estor Armesto
and Fabio Dom\'{\i}nguez}
\affiliation{Instituto Galego de F\'{\i}sica de Altas Enerx\'{\i}as IGFAE, Universidade de Santiago de Compostela, 15706 Santiago de Compostela, Galicia-Spain}
\author{Alex Kovner}
\affiliation{Physics Department, University of Connecticut, 2152 Hillside
Road, Storrs, CT 06269-3046, USA}

\author{Michael Lublinsky}
\affiliation{Physics Department, Ben-Gurion University of the Negev, Beer Sheva 84105, Israel}

\author{Vladimir V. Skokov}
\affiliation{Department of Physics, North Carolina State University, Raleigh, NC 27695, USA}
\affiliation{RIKEN/BNL Research Center, Brookhaven National
  Laboratory, Upton, NY 11973, USA}
  
\date{\today}

\begin{abstract}
We introduce the notion of the Color Glass Condensate (CGC) density matrix $\hat\rho$. This generalizes the concept of probability density for the distribution of the color charges in the hadronic wave function and is consistent with understanding the CGC as an effective theory after integration of part of the hadronic degrees of freedom.  We derive the evolution equations for the density matrix and show that the JIMWLK evolution equation arises here as the evolution of diagonal matrix elements of $\hat\rho$ in the color charge density basis.
We analyze  the behavior of this density matrix under high energy evolution and show that its purity decreases with energy. 
We show that the evolution equation for the density matrix has the celebrated Kossakowsky-Lindblad form describing the non-unitary evolution of the density matrix of an open system.   
Additionally,  we consider the dilute limit and demonstrate that, at large rapidity,  the entanglement entropy of the density matrix grows linearly with rapidity according
 to $\frac{d}{d y}S_e=  \gamma$, where $\gamma$ is the leading BFKL eigenvalue.
We also discuss 
the evolution of $\hat\rho$ in the saturated regime and relate it to the Levin-Tuchin law and find that the entropy again grows linearly with rapidity, but at a slower rate. 
By analyzing the dense and dilute regimes of the full density matrix we are able to establish a duality between the regimes. Finally we introduce the Wigner functional derived from this density matrix and discuss how it can be used to determine the distribution of color currents, which may be instrumental in understanding  dynamical features of QCD at high energy.

\end{abstract}
\maketitle
\section{Introduction}

High-energy hadronic collisions at RHIC and the LHC have demonstrated  an unexpected collective behavior in particle production. In particular,
 multiple observations of the structure of  final states in p-p and p-Pb collisions
at the LHC, see e.g. \cite{Khachatryan:2010nk,Chatrchyan:2013nka,Adam:2016iwf,Aaboud:2016jnr,ATLAS:2018jht},  indicate a very nontrivial dynamics that leads to a  correlated structure
between produced particles. 

The current understanding of the origin of these correlations is based on two concurrent pictures: the dominance of the collisions -- final state interactions, which in today's most popular incarnation is described in terms of transport  and hydrodynamics,
and the dominance of the correlations in the wave functions at the initial state and very early stage of the collision. The latter option is most commonly analyzed in the framework of the Color Glass Condensate (CGC), for a review see  e.g. Ref.~\cite{Iancu:2002xk,McLerran:2008uj,Gelis:2010nm,Kovchegov:2012mbw}. 

The CGC approach as applied to date  has one unappealing feature. It can accommodate correlations between partons either in coordinate space, or in momentum space, but it does not give one a handle to explore the correlations between the two. On the other hand, one can expect such correlations on general grounds.

The present work is motivated by the idea that one should be able to get some sense about the correlation between the profile of the charge density and the current density in the hadronic wave function at high energy. One may expect, for example, that if the scattering catches a configuration in the projectile wave function with large density or large density gradient, this configuration will naturally also have  a large current, since it does not want to stay static for a long time. The large current then may translate into relatively high momenta of the particles in the wave function. 
That way we may be able to relate on the level of the initial hadronic state the density (number) fluctuation in the configuration space and momentum distribution of particles.

To achieve this goal we have to expand the usual CGC vocabulary and introduce a novel concept in this field of studies -- the CGC density matrix. In the standard CGC approach 
one is only interested in the diagonal matrix elements of the density matrix $\hat \rho$ (usually denoted by $W$), which is sufficient for the calculation of a large number of observables. However, in
order to study the properties we alluded to before, the off-diagonal elements of the density matrix are required. We note that a very similar problem arose in
 two and many gluon production at different rapidities~\cite{Kovner:2006wr} (see also \cite{Iancu:2013uva}); in part, 
our current work will rely on intuition gained and guidance obtained  from Ref.~~\cite{Kovner:2006wr}.   

The plan of this paper is the following. In Sec. II we define the concept of the CGC density matrix $\hat\rho$. We stress that this is a completely different object than the one used to calculate the entropy of soft gluons for example in~\cite{Kovner:2015hga,Kovner:2018rbf}. In Sec. III we derive the evolution of this object with rapidity. The evolution equation turns out to be a natural generalization of the JIMWLK equation, and is of the Kossakowski-Lindblad form as could have been expected on general grounds.
In Sec. IV we consider the evolution of $\hat\rho$ in the weak and strong field regime in turn. We take a reasonable Gaussian ansatz for $\hat\rho$ and calculate the evolution of the parameters. We also calculate the entanglement entropy associated with $\hat\rho$ and show that is grows linearly with rapidity. The rate of growth is given by the leading BFKL exponent in the weak field case, and half of that in the saturation regime.
 In Sec. V we define the Wigner functional associated with $\hat\rho$. Finally we close in Sec. VI with a short discussion.
    
\section{The CGC density matrix}
Recall that the CGC is an ``effective field theory'' of high energy scattering. This is a somewhat loose term, but it does in fact have a fairly precise meaning given our 
standard CGC calculations. The standard practice in deriving the CGC wave function and the corresponding evolution equation is to integrate out all degrees of freedom in the hadronic wave function except for the integrated 
color charge density $j^a(x_\perp)=\int j^a(x_\perp,x^-) dx^-$. (Note that here and below we use $j = j^a t^a$ to denote the color charge densities instead of commonly used $\rho$. This change of notation is implemented to differentiate the color charge density and the density matrix.)  Thus CGC is in fact the effective field theory on the Hilbert space spanned by $j$.
 There is one subtlety here:  in general,  the components $j^a$ are not independent degrees of freedom, as they do not commute with each other. In the dense limit however the commutator between $j$'s can be neglected and they can be treated as independent.  We will follow this approach in the present paper  and will  treat  $j^a$ as commuting. 
 
Put in this way, our ambient CGC calculations are equivalent to simply integrating out a subset of quantum degrees of freedom in the hadronic wave function. 
Then it is natural to ask what is the reduced density matrix on the subspace of the Hilbert space spanned by the remaining  quantum degrees of freedom. In particular one can ask what is the 
entanglement entropy of this reduced density matrix, and how it evolves with energy (for indirectly related  calculations of the entanglement entropy in high energy collisions see Refs.~\cite{Elze:1994hj,Kutak:2011rb,Peschanski:2012cw,Peschanski:2016hgk,Berges:2017zws,Berges:2017hne,Hagiwara:2017uaz,Neill:2018uqw,Liu:2018gae}). These are the questions we will be gearing to ask in the present work.

One may wonder why it is that in all the CGC calculations to date~\footnote{Or at least in most calculations, with
the exception of Refs.~\cite{Hentschinski:2005er,Kovner:2006ge,Kovner:2006wr,Iancu:2013uva}. Although the full importance of the density matrix was not recognized in these papers at the time, the relation to the current work is actually very direct and will be discussed below.} there was no apparent need to define the full density matrix. 
The answer is that
the knowledge of the full density matrix is not always necessary. In particular, if one considers the calculation of observables which are functions of $j$ only, and not of their canonical conjugates, it is sufficient to know the diagonal matrix elements of $\hat\rho$ in the basis of charge density $j$.
Most of the observables considered  so far, like observables involving only the softest gluons in the CGC wave function, are of this type.

To elaborate on this further,  suppose the density matrix of the valence gluons in this particular basis is known and can be written as  (we suppress the color and coordinate indexes for simplicity)
\beq\label{jj'}
 \langle j|\hat\rho|j'\rangle\equiv \rho[j,j']\,.
\eeq
Now to include the soft gluons into consideration we find the soft gluon vacuum in the presence of the valence color charge,
$|s[j]\rangle$. The density matrix on the full (valence plus soft) Hilbert space is
\beq \hat\rho_{v+s}=|s[\hat j] \rangle\hat\rho\langle s[\hat j]|\,,
\eeq
where $\hat j$ is the color charge density operator.

Now suppose we need to calculate a matrix element of some operator which involves only the soft(est) gluon operators $\hat O(a,a^\dagger)$.
\beq {\rm Tr} [\hat O\hat\rho_{v+s}]= {\rm Tr} \Big[\hat O| s[\hat j]\rangle\hat\rho\langle s[\hat j]|\Big]= {\rm Tr} \Big[\langle s[\hat j]|\hat O|s[\hat j]\rangle\hat \rho \Big]=\int Dj \langle s[j]|\hat O|s[j]\rangle\rho[j,j]\,,
\label{Eq:Oj}
\eeq
where the last equality follows since the matrix element $\langle s[\hat j]|\hat O|s[\hat j]\rangle$ involves only the operator $\hat j$, and so in the eigenbasis $j$, we only require diagonal matrix elements of $\hat \rho$. As we alluded to before, 
we assume here that $j$ is large enough so that we can treat different color components of the charge density as mutually commuting. 

Equation \eqref{Eq:Oj} reduces to the formulae which were used in the CGC EFT  for averaging over the valence space with the weight functional 
\beq 
W[j]\equiv \rho[j,j]\,.
\eeq
We thus conclude that  indeed for the operators of this type we only need to know the diagonal matrix elements of $\hat \rho$. These diagonal matrix elements are encoded in the probability density functional $W[j]$ routinely used in the CGC approach.

However, this is not the only set of operators that are of interest in high energy scattering. One stark example of an operator of a different kind is the $S$ matrix for dense-dense scattering.
The eikonal $\hat S$-matrix for scattering on a strong color field acts on the color charge density operators by rotating them by the eikonal phase
\beq
\hat S^\dagger\hat j(x_\perp)\hat S=V(x_\perp)\hat j(x_\perp)\, ,
\eeq
where $V(x_\perp)$ is a unitary matrix. For strong fields $V(x_\perp)$ is an arbitrary element of the SU($N$) group and may be  arbitrarily far away from the unit matrix. It is therefore obvious that the $\hat S$-matrix is not a diagonal operator in the $j$ basis, and so non-diagonal matrix elements of the density matrix must be important in its evaluation.

Another example of this type of observable is the multi-gluon production probability where gluons are produced at different rapidities. Indeed,
when calculating multiple gluon production where the rapidities of gluons were significantly different~\cite{Kovner:2006wr}, see also Ref.~\cite{Iancu:2013uva}, 
from the target perspective it was necessary  to introduce novel weight functionals that depended on two different $j$'s (or related to them by
eikonal factors $S$, see below).  

It is obvious that  the knowledge of $W[j]$ is not sufficient to determine the complete density matrix. In particular, in the MV model~\cite{McLerran:1993ni,McLerran:1994vd}, $W[j]=\exp(-\frac{4j^2}{\mu^2})$ (in this paper, in order to simplify equations, we we deviate from the conventional normalization of $\mu^2$); this  weight functional could correspond to a variety of different density matrices with very contrasting
 properties. One example would be
\beq 
\rho[j,j']=\exp\left(-\frac{2j^2}{\mu^2}-\frac{2j'^2}{\mu^2}\right)\,.
\eeq
This density matrix has a factorized form, and evidently corresponds to a pure state on the reduced Hilbert space.
Obviously,  this  is not the only possibility. A priori any density matrix of the form
\begin{align}\label{gauss0}
	\rho[j,j']={\cal N}
	\exp\Bigg \{
	 \int d^2x_\perp d^2 y_\perp 
	{\rm tr_{\rm c}} 
     &\Bigg[- 
	\mu ^{-2} (x_\perp, y_\perp) (j(x_\perp) +j'(x_\perp)) 
	 (j(y_\perp) +j'(y_\perp)) 
	 \notag \\
	 &
	 -	 \ \lambda^{-2}(x_\perp, y_\perp) {(j(x_\perp) -j'(x_\perp)) 
	 (j(y_\perp) -j'(y_\perp)) 
	 }
 \notag \\
&+  i A(x_\perp, y_\perp) (j(x_\perp)  + j'(x_\perp))(j(y_\perp) -  j'(y_\perp) )\Bigg]\Bigg\}
\end{align}
with real functions $\mu,\ \lambda, A$ of variables $x_\perp$ and $y_\perp$ 
is an allowed density matrix inasmuch as it reduces to the MV model for diagonal elements and is Hermitian. 
There is only one restriction.  The  parameters $\mu,\ \lambda, A$ have to satisfy additional constraints  in order for $\hat \rho$ to have probabilistic interpretation, i.e. all eigenvalues have to be positive (the overall normalization can always be adjusted).  This is equivalent to the requirement that for any positive integer  $n$
\begin{equation}
	{\rm Tr} \,\hat \rho^{n} \ge 	{\rm Tr}\, \hat \rho^{n+1} \,. 
\end{equation}
For $n=1$ this yields
\begin{equation}
	{\rm det} [ \lambda^2] \le 	{\rm det} [ \mu^2 ].
\end{equation}
Here the determinants are in the transverse position space
\footnote{There are no constraints on function $A$. This is not surprising; as we will demonstrate later, function $A$ can be always rotated away from the definition of  the density matrix by performing a unitary transformation. Thus, $A$ never contributes
 to the operators of the form ${\rm Tr}\, \hat \rho^n$.  }. We also will often use the inverse functions, e.g. $\mu^{-2} (x_\perp,y_\perp)$, defined as 
 \begin{equation}
 \label{Eq:Inversion}
\int d^2 z \, \mu^{-2} (x_\perp,z_\perp)  \mu^{2} (z_\perp,y_\perp)  = \delta^{(2)}(x_\perp-y_\perp)\,. 	
\end{equation}
Equation~\eqref{gauss0} should be read following the definition 
in Eq.~\eqref{Eq:Inversion}. 

\section{High energy evolution of the CGC density matrix.
}

Given that $\hat\rho$ contains more information than $W[j]$, the natural first question is how does it  evolve to high energy? 
To start answering this question, we first point out that  previously, in Ref.~\cite{Kovner:2006ge}, the evolution of the density functional for two gluon production at two significantly different rapidities was derived. 
The problem at hand is very similar to that discussed in  Ref.~\cite{Kovner:2006ge}. 
Although at the time of writing of Ref.~\cite{Kovner:2006ge} the newly introduced weight functional was not interpreted as a density matrix, we will show below that the evolution derived in Ref.~\cite{Kovner:2006ge}  can indeed be mapped on to the evolution of the density matrix. 
The evolution in question is
\beq
\frac{d}{dy}\rho[j,j']=\int \frac{d^2z_\perp}{2\pi} \Big[Q_i^a[z_\perp ,j]+Q_i^a[z_\perp ,j']\Big]^2\rho[j,j']\,,
\label{Eq:PEV}
\eeq
where $Q$ is defined by 
\begin{align}
	\notag Q_i^a[z_\perp ,j]&=\frac{g}{2\pi}\int d^2x_\perp\frac{(x_\perp-z_\perp)_i}{(x_\perp-z_\perp)^2}\Big[S^{ab}(z_\perp)-S^{ab}(x_\perp)\Big]J^b_R(x_\perp) \\ &=\frac{g}{2\pi}\int d^2x_\perp\frac{(x_\perp-z_\perp)_i}{(x_\perp-z_\perp)^2}J^c_L(x_\perp)\Big[S^{cb}(x_\perp)S^{{\dagger} \, ba}(z_\perp)-\delta^{ca}\Big]
\end{align}
and
\beq
\label{Eq:JR}
J_R^a(x_\perp)=-{\rm tr}_{\rm c} \left\{S(x_\perp)T^a\frac{\delta}{\delta S^\dagger(x_\perp)}\right\}\,,\ \ 
J_L^a(x_\perp)=-{\rm tr}_{\rm c} \left\{T^aS(x_\perp)\frac{\delta}{\delta S^\dagger(x_\perp)}\right\}\,,
\eeq
where tr$_{\rm c}$ denotes the trace over color indexes.  
Here, as usual, $S$ is the eikonal phase matrix 
for scattering of a probe gluon on the wavefunction~\footnote{Notice that $S$ differs from $V$ introduced earlier: while one stands for
a scattering matrix of a probe gluon on a target, another one denotes the scattering matrix of a target gluon on the projectile.}.
The matrix $S$ is determined by the color charge density via
\begin{equation}
\frac{i}{g}\partial_i[S^\dagger\partial_iS]=j.
\end{equation}
In the dilute limit (small color charge density) we explicitly have
\begin{equation}
	S(x_\perp)  ={\cal P}  \exp \left[ i \int dx^- g\alpha(x_\perp, x^-)  \right]  
	\label{Eq:EikonalPhase}
\end{equation}
with 
\beq
\alpha(x_\perp)=  - \int_{y_\perp}\frac{1}{\partial^2}(x_\perp,y_\perp)j(y_\perp) \equiv  
\int \frac{d^2y_\perp}{4\pi}  \, \ln \frac{ 1 } {\Lambda^2 | x_\perp-y_\perp|^2  }  j(y_\perp)   \, .
\eeq
In the following we allow ourselves to denote the argument of the density matrix intermittently by  either $j$, $\alpha$ or $S$, as all these objects are algebraically related to each other.

Recall that the standard JIMWLK Hamiltonian is given in terms of $Q$'s as
\beq
H_{\rm JIMWLK}=\int  \frac{d^2z_\perp}{2\pi} Q^a_i[z_\perp,j] Q^a_i[z_\perp,j].
\eeq
Thus, Eq.~\eqref{Eq:PEV} generalizes JIMWLK evolution equation~\cite{JalilianMarian:1997dw,Kovner:2000pt,Kovner:1999bj,Weigert:2000gi,Iancu:2000hn,Iancu:2001ad,Ferreiro:2001qy} to the full density matrix.

\subsection{Derivation of high energy evolution} 
In order to derive this equation, we will follow the same main steps as in Ref.~\cite{Kovner:2006wr}. Our discussion will be in the framework of scattering of some dilute projectile on a dense  target.

Consider an observable $\hat{\cal O}$ which depends only on the projectile degrees of freedom. The projectile scatters on the target, and the observable is measured in the asymptotic state long
 time after the scattering has taken place. The total rapidity interval in the collision is $Y$.
 We assume that the target has been boosted to rapidity $Y_0$ and that the operator $\hat{\cal O}$ depends on degrees of freedom between the rapidity $Y_0$ and $Y$. 
In other words, the operator $\hat{\cal O}$ itself does not depend on the target degrees of freedom.   It is an operator in the projectile Hilbert space and as such defines an observable measured 
in the direction of the projectile. 

Let us define the following object 
\beq\label{O}
{\cal O}[S,\bar S]\,=\,\langle\,P_{Y-Y_0}|\,
 \,(1\,-\,\hat {S}^\dagger)\,\hat{\cal O}\,
\,(1\,-\,\hat {\bar S}) 
|P_{Y-Y_0}\,\rangle \,,
\eeq
where $|P_{Y-Y_0}\rangle$ is the wave function of the projectile.
This object is related to the 
 actual observable in the scattering process via 
\beq\label{average}
\langle \hat{\cal O}\rangle\,=\,\int_{S,\bar S}\langle S|\hat\rho_{Y_0}|\bar S\rangle\,{\cal O}[S,\bar S]\,,
\eeq 
where  $\hat\rho_{Y_0}$ is  the {
 target} density matrix we are interested in. Note that although the operator $\hat {\cal O}$ itself does not depend on the target degrees of freedom, the $S$-matrix factors do, so that the matrix element over the projectile wave function becomes an operator on the target Hilbert space. Also, it depends only on the integrated target degrees of freedom -- as required for our definition of the density matrix.  The variables $S$ and $\bar S$ in Eq.~(\ref{average})  are the analogs of $j$ and $j'$ in Eq.~(\ref{jj'}).

We now want to trace the evolution by an additional rapidity $\Delta y$ so that the extra rapidity moves the observable away from the target. 
This can be achieved by boosting the target by an additional rapidity $\Delta y$ relative to the lab frame, or, alternatively, 
boosting the projectile together with the observable $\hat {\cal O}$ by the same rapidity, so that $\hat{\cal O}$ remains at the fixed rapidity from the projectile.  
 It is straightforward to do the latter. After boosting the projectile we have
\beq\label{O2}
{\cal O}_{\Delta y}[S,\bar S]\,=\,\langle\,P_{Y-Y_0}|\,
 C^{\dagger}_{\Delta y}\,(1\,-\,\hat {S}^\dagger)\,
C_{\Delta y}\,\,\,\,\hat{\cal O}\,\,\,
C^{\dagger}_{\Delta y}\,
\,(1\,-\,\hat {\bar S})\,C_{\Delta y}\,
|P_{Y-Y_0}\,\rangle\,,
\eeq
where for small $\Delta y$ the coherent operator $C_{\Delta y}$ for dilute projectile
(see for example Ref.~\cite{Kovner:2005nq} for the complete definition of the  operator $C_y$)   can be expanded into a  power series 
\begin{eqnarray}\label{expand}
C_{ \Delta y}&=&
1\,+\,i\,\int d^2x_\perp\,b_i^a(x_\perp)\int_{\,\Lambda}^{e^{\Delta y}\,
\Lambda}{dk^+\over \pi^{1/2}| k^+|^{1/2}} 
\left[a^{ a}_i(k^+, x_\perp)\,+\,a^{\dagger a}_i(k^+, x_\perp))\right]\,-\,\nonumber \\
&-& \left(\int d^2x_\perp\,b_i^a(x_\perp)\int_{\,\Lambda}^{e^{y}\,\Lambda}
{dk^+\over \pi^{1/2}| k^+|^{1/2}} 
\left[a^{ a}_i(k^+, x_\perp)\,+\,a^{\dagger a}_i(k^+, x_\perp))\right]\right )^2
{+{\cal O}(b_i^3)}\,.
\end{eqnarray}
Here $b^a_i(x_\perp)$ is the Weizsacker-Williams  field of the projectile, 
\beq b^a_i(x_\perp)=g\int d^2z_\perp {(x_\perp-z_\perp)_i\over (x_\perp-z_\perp)^2} j^a_P(z_\perp )\eeq
and $j_P^a$ is the color charge  density of the dilute projectile (not to be confused with $j$ defined above which in the present context is the color charge density of the target).


The evolution equation for the operator is obtained from
\beq
{d {\cal O}[S,\bar S]\over d y}\,=\,
\lim_{\Delta y\rightarrow 0}\frac{{\cal O}_{\Delta y}[S,\bar S]\,-\,
{\cal O}[S,\bar S]}{  \Delta y} \,.
\eeq
We  remind the reader the following key identities valid for any multigluon state in the projectile Hilbert space (see Ref.~\cite{Kovner:2005jc}):
\begin{eqnarray}\label{iden}
&&j^a_P\,\hat {\bar S} \,| P\rangle\,=\,J_R^a[\bar S]\,\hat {\bar S}\,| P\rangle\,;
\ \ \ \  \ \ \ \ \ \ \ \
\hat {\bar S} \,j^a_P\,| P\rangle\,=\,J_L^a[\bar S]\,\hat {\bar S}\,| P\rangle\,;
\nonumber \\ 
&& \langle P |\,j^a_P\,\hat S^\dagger\,=\,J_L^a[S]\, \langle P |\,\hat S^\dagger\,;
\ \ \ \  \ \ \ \ \ \ \ \
\langle P|\,\hat S^\dagger\, j^a_P\,=\,J_R^a[S]\, \langle P |\,\hat S^\dagger\,.
\end{eqnarray}

By construction, the operator $\hat{\cal O}$ commutes with the soft gluon operators $a$ and $a^\dagger$, since it only involves degrees of freedom at rapidities between $Y_0+\Delta y$ and $Y+\Delta y$.
We can thus take the averages of all the soft gluon operators in the soft gluon vacuum, since the projectile state before boost was the vacuum for these modes.
Combining the expansion (\ref{expand}) and the identities (\ref{iden}) we obtain
\beq\label{evolo}
{d\over dy} {\cal O}[S,\bar S]\,=-\,H_3[S,\bar S]\,\,
 {\cal O}[S,\bar S]
\eeq 
with 
\beq H_3[S,\bar S]\,\equiv\,  \frac{d^2z_\perp}{2\pi}\,
\left[Q^a_i[z_\perp ,S]\,+\,Q^a_i[z_\perp ,\bar S]\right]^2 \,.
\eeq

Now recalling Eq.~(\ref{average}) we see that we can integrate the evolution kernel ``by parts'', so that the derivatives in the operators $Q$ act on the target density matrix. This results in the evolution equation for matrix elements of the density matrix
\beq\label{evolrho}
{d\over dY} \langle S|\hat\rho_Y|\bar S\rangle\,=-\,H_3[S,\bar S]\,\,
 \langle S|\hat \rho_Y|\bar S\rangle\,.
\eeq

Note that we can rewrite this in the operator form  by ``integrating by parts'' $Q[j']$ so
that it does not act on $\rho[j,j']$ but acts on the operator whose expectation value we are calculating.
Indeed let us consider the evolution of an arbitrary observable ${\cal O}(j,j')$, 
\beq
\frac{d}{dy} \langle  {\cal O}\rangle =\int D j D j' {\cal O}(j,j')  \int  \frac{d^2z_\perp}{2\pi}\Big[Q_i^a[z_\perp ,j]+Q_i^a[z_\perp ,j']\Big]^2\rho[j,j'] \,.
\eeq
Integrating by parts in the functional integral $j'$, we arrive at 
\begin{align}
\frac{d}{dy} \langle {\cal O}   \rangle =\int D j D j' \int  \frac{d^2z_\perp}{2\pi}  \Big\{& {\cal  O}(j,j') Q_i^a[z_\perp ,j]  Q_i^a[z_\perp ,j]  
-2\left(Q_i^a[z_\perp ,j'] {\cal O}(j,j')\right) 
Q_i^a[z_\perp ,j]
\notag \\ &+
\left(Q_i^a[z_\perp ,j']
Q_i^a[z_\perp ,j']
{\cal O}(j,j')\right) 
\Big\} \rho[j,j'] \,. 
\end{align}
Since the operator ${\cal O}$ is arbitrary, this is equivalent to the evolution of the density matrix operator in the form
\beq\label{ev}
\frac{d}{dy}\hat\rho=\int  \frac{d^2z_\perp}{2\pi}  \Big[\hat Q_i^a[z_\perp ],\Big[\hat Q_i^a[z_\perp ],\hat\rho\Big]\Big]\,,  
\eeq
where the operator  $\hat Q_i^a[z_\perp ]$ is defined in such a way that for an arbitrary ket $|\psi\rangle$
\begin{equation}
	\langle j|\hat Q_i^a[z_\perp ] |\psi\rangle =   Q_i^a[z_\perp ,j]   \langle j|\psi\rangle\,.  
\end{equation}
The evolution equation for the density matrix has the celebrated {\it Kossakowsky-Lindblad} form (see the original  papers in Ref.~\cite{KOSSAKOWSKI1972247,Lindblad1976} )  with $\int d^2z_\perp\Big[\hat Q_i^a[z_\perp ],\Big[\hat Q_i^a[z_\perp ],\hat\rho\Big]\Big]$  being the {\it Lindbladian} of the system 
with the so-called {\it jump} or {\it Lindblad} operator $\hat Q_i^a[z_\perp ]$.  The fact that the evolution has this form is not surprising. The Lindblad master equation (here without the unitary part of the evolution) is the most general form of the Markovian evolution preserving the trace and the positivity of the reduced density matrix. 
This equation is ubiquitous in various fields of physics whenever a description of an open system is attempted, see e.g. Ref.~\cite{Banks:1983by}.    The meaning of the jump operator in this context is the amplitude of the process in which the ``environment'' experiences a quantum jump to a different level. This is very natural in the context of the high energy evolution, as
 $Q_i^a[z_\perp ,j]$ is precisely an amplitude of emission of a soft gluon. Such emission process does indeed change the quantum state of the soft ``environment''.

We note one interesting feature of Eqs.~(\ref{ev}, \ref{evolrho}). Specifically concentrating on Eq.~(\ref{evolrho}) we see that diagonal matrix elements of $\hat\rho$ evolve independently of the nondiagonal ones. This is due to the property of the operator $H_3$ discussed in detail in Ref.~\cite{Kovner:2006wr},
\begin{equation}
H_3[S,\bar S]\, F[S,\bar S]|_{\bar S=S}=H_{\rm JIMWLK}[S]\, F[S,S]\,,
\end{equation}
valid for an arbitrary function $F$. Thus the diagonal matrix elements of $\hat\rho$ indeed evolve according to the standard JIMWLK equation.
 
\subsection{Entropy growth}
The Lindbladian does not have the form of the Hamiltonian evolution in quantum mechanics, since the time derivative of $\hat \rho$ is not given by a commutator with an Hermitian operator. Thus, the entropy of $\hat\rho$ increases in the course of the  evolution, as 
 well known, for the Lindblad master equation. 

Here, for completeness of the discussion,  we demonstrate this explicitly  in the following simple way. 
Let us examine the effect of the evolution on a pure state and  consider the evolution of $\hat \rho^2$. To reduce the notational clutter, in this section we will use the shorthand notation
$\rho_{jj'}=\rho[j,j']$ and 
$ \left( \vec{Q}_j^a\right)_i = Q_i^a[z_\perp ,j]$. We get 
\begin{align}
\frac{d}{dy}(\hat \rho^2)_{jj'}&=\int Dk\,\left(\frac{d\rho_{jk}}{dy}\rho_{kj'}+\rho_{jk}\frac{d\rho_{kj'}}{dy}\right) \nonumber \\
&=\int Dk \int \frac{d^2z_\perp}{2\pi}\left\{\left[(\vec{Q}_j^a+\vec{Q}_k^a)^2\rho_{jk}\right]\rho_{kj'}+\rho_{jk}\left[(\vec{Q}_k^a+\vec{Q}_{j'}^a)^2\rho_{kj'}\right]\right\} \nonumber \\
&=\int Dk \int \frac{d^2z_\perp}{2\pi}\left\{(\vec{Q}_j^a+\vec{Q}_{j'}^a)^2\rho_{jk}\rho_{kj'}-2\vec{Q}_j^a\vec{Q}_{j'}^a\rho_{jk}\rho_{kj'} \right. \nonumber \\
&\qquad\left.+\left[(2\vec{Q}_j^a\vec{Q}_k^a+\vec{Q}_k^{a2})\rho_{jk}\right]\rho_{kj'}+\rho_{jk}\left[(\vec{Q}_k^{a2}+2\vec{Q}_k^a\vec{Q}_{j'}^a)\rho_{kj'}\right]\right\}.
\label{Eq:rho2}
\end{align}
In the first term of the equality \eqref{Eq:rho2}  the $k$ integration is trivial. Using the pure state condition $\hat \rho^2=\hat \rho$ one recognizes in this term the derivative of $\hat \rho$. The rest of the terms can be rearranged after using integration by parts on the jump operators  $\vec{Q}_k^a$:
\begin{align}\label{evolrho2}
	\frac{d}{dy}(\hat \rho^2)_{jj'}&=\frac{d}{dy}\rho_{jj'}-2\int Dk\int\frac{d^2z_\perp}{2\pi}\left[(\vec{Q}_j^a+\vec{Q}_k^a)\rho_{jk}\right]\left[(\vec{Q}_k^a+\vec{Q}_{j'}^a)\rho_{kj'}\right].
\end{align}
It is clear that the evolution of $\hat \rho$ and $\hat \rho^2$ differs by a non-trivial term, indicating that a pure state becomes mixed after evolution. One can go one step further and take the trace of Eq.~(\ref{evolrho2}). Taking into account that the trace of $\hat \rho$ is always 1, we get
\begin{equation}
	\frac{d}{dy}\text{Tr} \, \hat \rho^2=-2\int Dj\,Dk\int\frac{d^2z_\perp}{2\pi}\left[(\vec{Q}_j^a+\vec{Q}_k^a)\rho_{jk}\right]\left[(\vec{Q}_k^a+\vec{Q}_j^a)\rho_{kj}\right]\, .
\end{equation}
Given that $\hat\rho$ is Hermitian and $Q$ is real, the integrand is clearly positive definite. Therefore we conclude that
\begin{equation}
\frac{d}{dy}\text{Tr}\, \hat \rho^2<0 \,. 
\label{Eq:neg}
\end{equation}
Thus the evolution changes  the density matrix  $\hat\rho$ such that it does not correspond to a pure state anymore. In particular, by using the standard definition of the  Renyi entropy  $S_R=-\ln \text{Tr}\hat\rho^2$, we find
\beq 
	\frac{d}{dy} S_R>0,
\eeq
showing that the entropy of the density matrix increases due to evolution. We will return to an explicit calculation of the Renyi entropy in the following section. 
The property  of decoherence and \eqref{Eq:neg} are well-known in the context of the Lindblad master equation, but has been derived in the CGC framework for the first time here. 

\section{Evolution in Gaussian approximation}
In the previous section we derived the evolution of the density matrix. 
It is highly nonlinear and non-local due to the complexity of the jump operator $Q$.  To get some idea on how the evolution affects the off diagonal matrix elements of $\hat \rho$, we will follow the ideas introduced before for the diagonal components of the density matrix in the context of the JIMWLK evolution equation, see e.g. Refs.~\cite{Iancu:2002aq,Iancu:2011nj}. Namely we will consider a Gaussian approximation for the density matrix, and will derive the evolution equations for the effective parameters. We consider the following approximation to $\hat\rho$:
\begin{align}\label{gauss}
	\rho[\alpha,\alpha']={\cal N}
	\exp\Bigg \{
	 \int d^2x_\perp d^2 y_\perp 
	{\rm tr_{\rm c}} 
     &\Bigg[- 
	 (\alpha(x_\perp) +\alpha'(x_\perp)) \mu_y ^{-2}(x_\perp, y_\perp) 
	 (\alpha(y_\perp) +\alpha'(y_\perp)) 
	 \notag \\
	 &
	 -	  \, (\alpha(x_\perp) -\alpha'(x_\perp)) \lambda_y^{-2}(x_\perp, y_\perp) 
	 (\alpha(y_\perp) -\alpha'(y_\perp)) 
 \notag \\
&+  i(\alpha(x_\perp)  + \alpha'(x_\perp))A_y(x_\perp, y_\perp)\, (\alpha (y_\perp) -  \alpha'(y_\perp) )\Bigg] 
\Bigg\}\,.
\end{align}
Here, to simplify the derivation,  we introduced the field $\alpha$ instead of the color charge density. We anticipate that the rapidity evolution of the density matrix  will be encoded in  the rapidity dependence of  parameters $\mu_y^2$, $\lambda_y$ and $A_y$. In principle $\mu_y^2$, $\lambda_y$ and $A_y$ can be taken as arbitrary matrices in color space, and the form Eq.~(\ref{gauss})
can accommodate such a general choice. However, color neutrality requires all these matrices to be proportional to identity, and we will restrict the general ansatz correspondingly. 

We start by deriving the evolution of these parameters in the dilute regime; we then also consider the approach to the  saturated regime.  

\subsection{Gaussian approximation for density matrix evolution in the dilute regime}

To derive the evolution for the three parameters in Eq.~\eqref{gauss}  
we have to  consider three different averages $\langle \hat O_i \rangle$ and require that their evolution is reproduced by the Gaussian ansatz.
The natural choice is to take the averages of the three simple linearly independent operators: 
\beq
\hat O_i=\left \{\alpha^a(x_{1 \perp})  \frac{\delta}{ \delta {\alpha}^{ a}(x_{2 \perp})} ,\ \ \alpha^a(x_{1 \perp})  {\alpha}^{a}(x_{2 \perp})\ \ ,  \frac{\delta}{ \delta \alpha^{a}(x_{1 \perp})} \frac{  \delta}{\delta {\alpha}^{ a}(x_{2 \perp})}\right\}\,.
\eeq
For each operator in this set  we first calculate the corresponding expectation value 
\beq
\langle \hat O_i\rangle_{(\mu_y,\lambda_y,A_y)}\equiv \text{Tr}[\hat O_i\hat\rho]
\eeq
and then take its derivative with respect to rapidity
\beq
\frac{d}{dy}\langle \hat O_i\rangle =\frac{\partial \langle \hat O_i\rangle}{\partial\mu_y}\frac{d\mu_y}{dy}+\frac{\partial \langle \hat 
O_i\rangle }{\partial\lambda_y}\frac{d\lambda_y}{dy}+\frac{\partial \langle \hat O_i\rangle}{\partial A_y}\frac{dA_y}{dy} \,.
\label{Eq:der}
\eeq
Since the evolution of each expectation value is dictated by Eq.~\eqref{Eq:PEV}, Eq.~\eqref{Eq:der}
 has to be equated to 
 \begin{align}
\text{Tr}[\hat O_i\frac{d}{dy}\hat\rho]&=\int D\alpha \left\{\hat O_i(\alpha',\alpha)\int d^2z_\perp\left[Q^a_k(z_\perp, \alpha)+Q^a_k(z_\perp, \alpha')\right]^2\rho(\alpha,\alpha')\right\}_{\alpha'=\alpha} \notag \\ &=\int d^2z_\perp\text{Tr}\left\{\hat O_i[\hat Q_i^a(z_\perp),[\hat Q_i^a(z_\perp),\hat\rho]]\right\} \,.
\end{align}
In fact, given our choice of operators $\hat O_i$ it is easier to rewrite the last expression using the cyclic property of the trace as
\beq
\text{Tr}[\hat O_i\frac{d}{dy}\hat\rho]=\int  \frac{d^2z_\perp}{2\pi} \text{Tr}\left\{\hat\rho[\hat Q_i^a(z_\perp),[\hat Q_i^a(z_\perp),\hat O_i]]\right\} \,.
\eeq
Thus the equations that determine the evolution of the parameters of the density matrix in the Gaussian approximation are
\beq
\frac{\partial \langle \hat O_i\rangle}{\partial\mu_y}\frac{d\mu_y}{dy}+\frac{\partial \langle \hat O_i\rangle }{\partial\lambda_y}\frac{d\lambda_y}{dy}+\frac{\partial \langle \hat O_i\rangle }{\partial A_y}\frac{dA_y}{dy}=\int D\alpha \left\{O_i(\alpha',\alpha)\int \frac{d^2z_\perp}{2\pi} \left[Q^a_k(z_\perp, \alpha)+Q^a_k(z_\perp, \alpha')\right]^2\rho(\alpha,\alpha')\right\}_{\alpha'=\alpha} \,.
\eeq

Proceeding with the plan outlined above,  we calculate the averages in the Gaussian state density matrix:
\begin{eqnarray}
\left \langle \alpha^a(x_{1 \perp})\alpha^a(x_{2 \perp})\right \rangle&=&
\frac{(N_c^2-1)}{8}\mu_y^2(x_{1 \perp},x_{2 \perp})\,, \\
\left \langle \frac{\delta}{\delta\alpha^a(x_{1 \perp})}\frac{\delta}{\delta\alpha^a(x_{2 \perp})} \right \rangle&=&-2(N_c^2-1)\left[\lambda_y ^{-2}(x_{1 \perp},x_{2 \perp})-\frac{1}{4}\int d^2y_\perp d^2y_\perp'A(x_{1 \perp},y_\perp)\mu_y^2(y_\perp, y_{1 \perp})A_y(y',x_{2 \perp})\right]\,,\\
\left  \langle\frac{\delta}{\delta\alpha^a(x_{1 \perp})}\alpha^a(x_{2 \perp}) \right \rangle&=&\frac{(N_c^2-1)}{4}\ i\int d^2y_\perp A_y(x_{1 \perp},y_\perp)\mu_y^{2}(y_\perp, x_{2 \perp})\, ,
\end{eqnarray}
where we have explicitly assumed that $x_{1 \perp}\ne x_{2 \perp}$. One has to be more careful with the derivation for $x_{1 \perp}= x_{2 \perp}$.

In the dilute limit, operator $\hat Q$ can be expanded to leading order in $\alpha$:
\begin{equation}
	Q_i^a[z_\perp ,\alpha] \approx  
	-\frac{g^2}{2\pi} 
	\int d^2x_\perp 
	\frac{(x_\perp-z_\perp)_i}{(x_\perp-z_\perp)^2} 
	T^{d}_{ab} \left( \alpha^d (z_\perp) -  \alpha^d (x_\perp) \right) \frac{\delta}{\delta \alpha^b(x_\perp)} \,,
\end{equation}
with corrections  of order $\alpha^2$. 
In this limit, we also find
\begin{eqnarray}
&& [\hat Q_i^a(z_\perp),[\hat Q_i^a(z_\perp),\alpha^f(x_{1 \perp})\alpha^f(x_{2 \perp})]]
\notag \\ &&\approx  N_c \left( \frac{g^2}{2\pi} \right)^2 \int d^2z_\perp   \Bigg\{  -\frac{ (x_{1 \perp}-x_{2 \perp})^2 } {(x_{1 \perp}-z_\perp) ^2 (x_{2 \perp}-z_\perp) ^2 }
	\left(\alpha^d(z_\perp)-\alpha^d(x_{2 \perp}) \right) \left(\alpha^d(z_\perp)-\alpha^d(x_{1 \perp}) \right)
\\
&-&\frac{1}{(x_{2 \perp}-z_\perp) ^2}\alpha^a(z_\perp)\alpha^a(x_{2 \perp})-\frac{1}{(x_{1 \perp}-z_\perp) ^2}\alpha^a(z_\perp)\alpha^a(x_{1 \perp})+\left[\frac{1}{(x_{2 \perp}-z_\perp) ^2}+\frac{1}{(x_{1 \perp}-z_\perp) ^2}\right]\alpha^a(z_\perp)\alpha^a(z_\perp)\Bigg\}\nonumber, 
\end{eqnarray}
\begin{eqnarray}
&&[[\frac{\delta}{\delta \alpha^f(x_{1 \perp})}\frac{\delta}{\delta\alpha^f(x_{2 \perp})},\int d^2z_\perp\,  \hat Q^a_i(z_\perp)], \hat Q^a_i(z_\perp)]
\notag \\&&\approx
N_c\frac{g^4}{4\pi^2}\Bigg\{ \delta^{(2)}(x_{1 \perp}-x_{2 \perp})\int d^2x_\perp d^2y_\perp\left[\frac{1}{(x_\perp- x_{1 \perp})^2}+\frac{1}{(y-x_{1 \perp})^2}-\frac{(x_\perp-y_\perp)^2}{(x_\perp- x_{1 \perp})^2(y-x_{1 \perp})^2}\right] \notag \\ &\times&  \frac{\delta}{\delta\alpha^a(x_\perp)}\frac{\delta}{\delta\alpha^a(y_\perp)}\nonumber\\
&-&\int d^2z_\perp\frac{(x_{2 \perp}-x_{1 \perp})^2}{(x_{2 \perp}-z_\perp) ^2(x_{1 \perp}-z_\perp) ^2}\frac{\delta}{\delta\alpha^a(x_{1 \perp})}\frac{\delta}{\delta\alpha^a(x_{2 \perp})}\nonumber\\
&+& \int d^2z_\perp\left[\frac{(z_\perp- x_{1 \perp})^2}{(z_\perp- x_{2 \perp})^2(x_{1 \perp}-x_{2 \perp})^2}-\frac{1}{(x_{1 \perp}-x_{2 \perp})^2}\right]\frac{\delta}{\delta\alpha^a(z_\perp)}\frac{\delta}{\delta\alpha^a(x_{1 \perp})}\\
&+&\int d^2z_\perp\left[  \frac{(z_\perp- x_{2 \perp})^2}{(z_\perp- x_{1 \perp})^2(x_{2 \perp}-x_{1 \perp})^2}-\frac{1}{(x_{1 \perp}-x_{2 \perp})^2}   \right]\frac{\delta}{\delta\alpha^a(z_\perp)}\frac{\delta}{\delta\alpha^a(x_{2 \perp})}\Bigg \}\nonumber
\end{eqnarray}
and, finally,
\begin{eqnarray}
\label{Eq:Aevo}
&&[[\frac{\delta}{\delta\alpha^f(x_{1 \perp})}\alpha^f(x_{2 \perp}),\int d^2z_\perp\, \hat Q^a_i(z_\perp)],\hat Q^a_i(z_\perp)]
\notag \\ &&\approx N_c\frac{g^2}{4\pi^2}\Bigg\{ \int d^2x_\perp\ 2\frac{(x_\perp- x_{1 \perp})\cdot(x_{2 \perp}-x_{1 \perp})}{(x_\perp- x_{1 \perp})^2(x_{2 \perp}-x_{1 \perp})^2}\frac{\delta}{\delta\alpha^a(x_\perp)}\alpha^a(x_{1 \perp})\\
&+&\int d^2x_\perp\left[-2\frac{(x_\perp- x_{1 \perp})\cdot(x_{2 \perp}-x_{1 \perp})}{(x_\perp- x_{1 \perp})^2(x_{2 \perp}-x_{1 \perp})^2}+\frac{1}{(x_\perp- x_{1 \perp})^2}\right]\frac{\delta}{\delta\alpha^a(x_\perp)}\alpha^a(x_{2 \perp})\nonumber\\
&+&\int d^2z_\perp\left[-2\frac{(x_{2 \perp}-z_\perp) \cdot(x_{1 \perp}-z_\perp) }{(x_{2 \perp}-z_\perp) ^2(x_{1 \perp}-z_\perp) ^2}+\frac{1}{(x_{2 \perp}-z_\perp) ^2}\right]\frac{\delta}{\delta\alpha^a(x_{1 \perp})}\alpha^a(z_\perp)\nonumber\\
&+&\int d^2z_\perp\left[2\frac{(x_{2 \perp}-z_\perp) \cdot(x_{1 \perp}-z_\perp) }{(x_{2 \perp}-z_\perp) ^2(x_{1 \perp}-z_\perp) ^2}-\frac{1}{(x_{2 \perp}-z_\perp) ^2}-\frac{1}{(x_{1 \perp}-z_\perp) ^2}\right]\frac{\delta}{\delta\alpha^a(x_{1 \perp})}\alpha^a(x_{2 \perp}) \Bigg\}\nonumber \,.
\end{eqnarray}
In what follows, in order to simplify the derivation, we set $A=0$. As an  initial condition, $A=0$ is preserved by the evolution 
and is thus setting $A=0$ at any rapidity is compatible with Eq.~\eqref{Eq:Aevo}. As we will discuss in the next subsection, the function $A$ in general does not affect the eigenvalues of the density matrix 
and thus can be set equal to zero for the purpose of evaluating the entanglement entropy. For a general operator, the evolution of the 
function $A$, however, can be important and should not be omitted.  

Note that the evolution for $\mu_y^2$ and $\lambda_y^2$ decouple in the Gaussian approximation with $A=0$, and the evolution equations become:
\begin{eqnarray}\label{mueq}
\frac{\partial}{\partial y}\mu_y^2(x_{1 \perp},x_{2 \perp})&=& \frac{N_c}{2\pi}  \left( \frac{g^2}{2\pi} \right)^2 \int d^2z_\perp   \Bigg\{  -\frac{ (x_{1 \perp}-x_{2 \perp})^2 } {(x_{1 \perp}-z_\perp) ^2 (x_{2 \perp}-z_\perp) ^2 }
\\	& \times& \left(\mu_y^2(x_{1 \perp},x_{2 \perp})+\mu_y^2(z_\perp, z_\perp)-\mu_y^2(z_\perp, x_{2 \perp})-\mu_y^2(x_{1 \perp},z_\perp)\right)
\nonumber\\
&-&\frac{1}{(x_{2 \perp}-z_\perp) ^2}\mu_y^2(z_\perp, x_{2 \perp})-\frac{1}{(x_{1 \perp}-z_\perp) ^2}\mu_y^2(x_{1 \perp},z_\perp) \notag \\ &+&\left[\frac{1}{(x_{2 \perp}-z_\perp) ^2}+\frac{1}{(x_{1 \perp}-z_\perp) ^2}\right]\mu_y^2(z_\perp, z_\perp)\Bigg\}
\notag
\end{eqnarray}
and 
\begin{eqnarray}\label{lambdaeq}
&&\frac{\partial}{\partial y}\lambda_y^{-2}(x_{1 \perp},x_{2 \perp})=
\frac{N_c}{2\pi} \frac{g^4}{4\pi^2}\Bigg\{ -\delta^{(2)}(x_{1 \perp}-x_{2 \perp})\int d^2x_\perp d^2y_\perp\frac{(x_\perp-y_\perp)^2}{(x_\perp- x_{1 \perp})^2(y-x_{1 \perp})^2}\lambda_y^{-2}(x_\perp,y_\perp)\\
&+&\int d^2z_\perp\left[\frac{(z_\perp- x_{1 \perp})^2}{(z_\perp- x_{2 \perp})^2(x_{1 \perp}-x_{2 \perp})^2}\lambda_y^{-2}(z_\perp, x_{1 \perp})+ \frac{(z_\perp- x_{2 \perp})^2}{(z_\perp- x_{1 \perp})^2(x_{2 \perp}-x_{1 \perp})^2}\lambda_y^{-2}(z_\perp, x_{2 \perp})\right.\nonumber\\
&&\hskip 1.2cm \left.-\frac{(x_{2 \perp}-x_{1 \perp})^2}{(x_{2 \perp}-z_\perp) ^2(x_{1 \perp}-z_\perp) ^2}\lambda_y^{-2}(x_{1 \perp},x_{2 \perp})\right]\nonumber\\
&-& \int d^2z_\perp\left[\frac{1}{(x_{1 \perp}-x_{2 \perp})^2}\lambda_y^{-2}(z_\perp, x_{1 \perp})+\frac{1}{(x_{1 \perp}-x_{2 \perp})^2}   \right]\lambda_y^{-2}(z_\perp, x_{2 \perp})\nonumber \\
&+&\delta^{(2)}(x_{1 \perp}-x_{2 \perp})\int d^2x_\perp d^2y_\perp\left[\frac{1}{(x_\perp- x_{1 \perp})^2}+\frac{1}{(y-x_{1 \perp})^2}\right]\lambda_y^{-2}(x_\perp,y_\perp)\Bigg\} \nonumber \,.
\end{eqnarray}

These equations can be further simplified. Concentrating first on Eq.~(\ref{mueq}), we note that it is convenient to define
\begin{equation}
\bar\mu_y^2(x_\perp,y_\perp)\equiv \mu_y^2(x_\perp,y_\perp)-\frac{1}{2}\mu_y^2(x_\perp,x_\perp)-\frac{1}{2}\mu_y^2(y_\perp,y_\perp) \,.
\end{equation}
The evolution equation for this quantity becomes
\begin{align}
\frac{\partial}{\partial y}\bar\mu_y^2(x_{1 \perp},x_{2 \perp})=&\frac{N_c}{2\pi}  \left( \frac{g^2}{2\pi} \right)^2 \int d^2z_\perp   \Bigg\{  -\frac{ (x_{1 \perp}-x_{2 \perp})^2 } {(x_{1 \perp}-z_\perp) ^2 (x_{2 \perp}-z_\perp) ^2 } \notag \\  &\times
	\left[\bar\mu_y^2(x_{1 \perp},x_{2 \perp})+\bar\mu_y^2(z_\perp, z_\perp) -\bar\mu_y^2(z_\perp, x_{2 \perp})-\bar\mu_y^2(x_{1 \perp},z_\perp)\right]\Bigg\} \,.
	\label{mueq1}
\end{align}
Physically $\bar\mu^2$ differs from $\mu^2$ only by terms that do not depend on one of the coordinates. In momentum space $\bar\mu_y^2$ and $\mu_y^2$ are therefore identical except possibly for zero momentum modes. For this reason we will not distinguish between $\bar\mu_y^2$ and $\mu_y^2$ in the following.

As for Eq.~(\ref{lambdaeq}) we note that the last two terms in this equation are proportional to zero momentum modes of $\lambda_y^{-2}$. Thus if $\int d^2x_\perp\lambda_y^{-2}(x_\perp,y_\perp)=0$ these terms drop out. It is also easily verified that this condition is preserved by Eq.~(\ref{lambdaeq}), i.e. assuming $\int d^2x_\perp\lambda^{-2}_{y_0}(x_\perp,y_\perp)=0$ at initial rapidity $y_0$ one has $\int d^2x_\perp\lambda^{-2}_y(x_\perp,y_\perp)=0$ at any rapidity $y$. We will thus drop these terms and simplify Eq.~(\ref{lambdaeq}) to
\begin{eqnarray}\label{lambdaeq1}
&&\frac{\partial}{\partial y}\lambda_y^{-2}(x_{1 \perp},x_{2 \perp})=
\frac{N_c}{2\pi} \left( \frac{g^2}{2\pi} \right)^2\Bigg\{-\delta^{(2)}(x_{1 \perp}-x_{2 \perp})\int d^2x_\perp d^2y_\perp\frac{(x_\perp-y_\perp)^2}{(x_\perp- x_{1 \perp})^2(y-x_{1 \perp})^2}\lambda_y^{-2}(x_\perp,y_\perp)\\
&+&\int d^2z_\perp\left[\frac{(z_\perp- x_{1 \perp})^2}{(z_\perp- x_{2 \perp})^2(x_{1 \perp}-x_{2 \perp})^2}\lambda_y^{-2}(z_\perp, x_{1 \perp})+ \frac{(z_\perp- x_{2 \perp})^2}{(z_\perp- x_{1 \perp})^2(x_{2 \perp}-x_{1 \perp})^2}\lambda_y^{-2}(z_\perp, x_{2 \perp}) \right.\notag \\ && \left. -   \frac{(x_{2 \perp}-x_{1 \perp})^2}{(x_{2 \perp}-z_\perp) ^2(x_{1 \perp}-z_\perp) ^2}\lambda_y^{-2}(x_{1 \perp},x_{2 \perp})\right]\Bigg\}\nonumber \,.
\end{eqnarray}

We observe that both Eq.~(\ref{mueq1}) and Eq.~(\ref{lambdaeq1}) are equivalent to different forms of the celebrated BFKL equation~\cite{Kuraev:1976ge,Fadin:1975cb}. 
Eq.~(\ref{mueq1}) is identical to the BFKL equation for scattering amplitudes while Eq.~(\ref{lambdaeq1}) is the BFKL equation for the correlator of the color charge density in the hadronic wave function. Thus, at high energy both $\mu_y^2$ and $\lambda_y^{-2}$ grow with the same leading BFKL exponential and  we have 
\begin{equation}\label{growth}
\mu_y^2\propto \exp (\gamma y),\ \ \ \ \lambda_y^2\propto \exp(-\gamma y),
\end{equation}
where 
\begin{equation}\label{gamma}
\gamma=\frac{4 \alpha_sN_c}{\pi}\,\ln 2\,.
\end{equation}

\subsection{High energy evolution, von Neumann entropy and decoherence in the dilute regime}
We already showed that the density matrix describing a pure initial state decoheres with evolution. Now our goal is to understand how this decoherence happens at high energies. 
The measure of such decoherence is the entanglement entropy. For the Gaussian density matrix,  it can be calculated. Let us start with the $N$-th Renyi entropy, which is somewhat easier to calculate than the von Neumann one.

As we alluded to before the parameter $A$ does not enter to the expression for the entropy. 
To prove this we note that in the definition of the density matrix it appears as part of a unitary basis change. In particular the density matrix $\hat\rho$ of Eq.~(\ref{gauss}) can be written as
\begin{equation}
\hat\rho=U\, \hat\rho' \,  U^\dagger \,,
\end{equation}
where
\begin{align}\label{gauss1}
	\rho'[\alpha,\alpha']=
	{\cal N} \exp\Bigg \{
	 \int d^2x_\perp d^2 y_\perp 
	{\rm tr_{\rm c}} 
     &\Bigg[- 
	(\alpha(x_\perp) +\alpha'(x_\perp))  \mu_y ^{-2}(x_\perp, y_\perp)\,
	 (\alpha(y_\perp) +\alpha'(y_\perp)) 
	 \notag \\
	 &
	 -	 {(\alpha(x_\perp) -\alpha'(x_\perp)) \lambda_y^{-2}(x_\perp, y_\perp) 
	 (\alpha(y_\perp) -\alpha'(y_\perp)) 
	 }
 \Bigg]\Bigg\}
\end{align}
and 
\begin{equation}
	U=\exp\left[i \, {\rm tr}_c 
	\int d^2x_\perp d^2y_\perp\, \alpha(x_\perp)A(x_\perp,y_\perp)\alpha(y_\perp)\right]\,.
\end{equation}
Thus $A$ does not affect the eigenvalues of $\hat\rho$, and  does not change the evolution of any operator of  the form ${\rm tr} \hat \rho^n$. For that reason, in the following we will set $A$ to zero.

Using the parametrization of the density matrix we can find the $N$-th Renyi entropy
\begin{equation}
	S_N = \frac{1}{1-N} 
	\ln \left[ {\rm Tr}  \left( \hat\rho \right)^N   \right] 
	\label{Eq:RenN}
 \end{equation}
 following the same steps as in  Ref.~\cite{Kovner:2015hga}. The trace of the density matrix to the $N$-th power is
 \begin{align}
	  {\rm Tr}  \left( \hat\rho \right)^N   & = 
	  {\cal N}^N
	  \int \prod_{i=1}^{N} D\alpha_i 
	  \exp \Big\{
		  \sum_{j=1}^{N}
	   \int d^2x_\perp d^2 y_\perp 
	 {\rm tr}_c 
	  \left[ 
	  -  \left( \alpha_j(x_\perp) + \alpha_{j+1}(x_\perp) \right) \mu^{-2}_y (x_\perp, y_\perp) 
	  \left( \alpha_j(y_\perp) + \alpha_{j+1} (y_\perp) \right)  
\right.  \notag  \\ &
\left. 
-	{(\alpha_j(x_\perp) -\alpha_{j+1}(x_\perp)) \lambda_y^{-2}(x_\perp, y_\perp)  
(\alpha_j(y_\perp) -\alpha_{j+1}(y_\perp)) 
	 }
	  \right] 
	  \Big\},
	 \label{Eq:TrrhoN}
 \end{align}
 with periodic boundary conditions in the replica space $\alpha_{N+1} = \alpha_{1}$. 
This integral is not diagonal in $\alpha$. The easiest way to proceed with integration over replicas is to transform 
the  expression in the exponential 
into  the Fourier replica space. It is  introduced according to  
\begin{equation}
	\alpha_j (x_\perp) 
	= \sum_{J = 1}^{N} 
	\tilde \alpha_J (x_\perp) e^{i \frac{2\pi}{ N} j J} \,. 
\end{equation}
This transformation satisfies the periodicity conditions  $\alpha_{j+N} = \alpha_{j}$. 
The reality of $\alpha_j (x_\perp)$ also leads to the relation  
\begin{equation}
	\tilde \alpha^*_J  = \tilde \alpha_{N-J}.  
\end{equation}
Let us consider two types of expressions  that we encounter in the calculation:  
\begin{equation}
	\sum_{j=1}^{N}
	\alpha_j(x_\perp) 
	\alpha_j(y_\perp)
	 = 
	 N 
	 \sum^N_{J=1} \tilde \alpha_J(x_\perp) \tilde \alpha^*_J(y_\perp) 
\end{equation}
and 
\begin{equation}
	\sum_{j=1}^{N}
	\alpha_j(x_\perp) 
	\alpha_{j+1}(y_\perp)
	 = 
	 N 
	 \sum^N_{J=1} \tilde \alpha_J(x_\perp) \tilde \alpha^*_J(y_\perp) 
	 e^{-i \frac{2\pi}{N} J}.  
\end{equation}
Using these we get 
 \begin{align}
	 \notag  {\rm Tr}  \left( \hat\rho \right)^N   & = 
	 	  {\cal N}^N
	  \int \prod_{I=1}^{N} D\tilde \alpha_i 
	  \exp \left\{
		  N
		  \sum_{J=1}^{N}
	   \int d^2x_\perp d^2 y_\perp 
	  {\rm tr}_c 
	  \left( 
	  - 2 \left[  \lambda_y^{-2}(x_\perp, y_\perp)  
	  +
	      \mu_y^{-2}(x_\perp, y_\perp)  
   \right] 
	  \tilde \alpha_J(x_\perp) \tilde \alpha^*_J(y_\perp)
\right. \right.
	  \\
& \left. \left. 
	  - 2 \left[ - {\lambda_y^{-2}(x_\perp, y_\perp) } 
	  +
	    \mu_y^{-2}(x_\perp, y_\perp) 
   \right] 
	  \tilde \alpha_J(x_\perp) \tilde \alpha^*_J(y_\perp) 
	  \cos \left( \frac{2\pi}{N} J\right)
	  \right) 
	  \right\} \,.
	 \label{Eq:TrrhoN1}
 \end{align}
The integration over $\tilde \alpha_N$ does not involve any factors of $\lambda$. This is an integration with respect 
to the center of mass in the replica space. This integral will be canceled by the normalization of the density matrix.
Thus we have 
\begin{align}
	 \notag  {\rm Tr}  \left( \hat\rho \right)^N   & = 
	  {\cal N}^{N-1}
	  \int \prod_{I=1}^{N-1} D\tilde \alpha_i 
	  \exp \left\{
		  N
		  \sum_{J=1}^{N-1}
	   \int d^2x_\perp d^2 y_\perp 
	  {\rm tr}_c 
	  \left( 
	  - 2 \left[ {\lambda_y^{-2}(x_\perp, y_\perp) } 
	  +
	      \mu_y^{-2}(x_\perp, y_\perp) 
   \right] 
	  \tilde \alpha_J(x_\perp) \tilde \alpha^*_J(y_\perp)
\right. \right.
	  \\
& \left. \left. 
	  - 2 \left[ - {\lambda_y^{-2}(x_\perp, y_\perp) } 
	  +
	    \mu_y^{-2}(x_\perp, y_\perp)  
   \right] 
	  \tilde \alpha_J(x_\perp) \tilde \alpha^*_J(y_\perp) 
	  \cos \left( \frac{2\pi}{N} J\right)
	  \right) 
	  \right\}
	  \notag \\ & =
	 \prod_{I=1}^{N-1}
	  \det 
	  \left[ 
		  \frac12	  \left(  { \mu_y^2  \lambda_y^{-2}} 
	  +
  	1  
  \right)
+ 
	  \frac12 \left(  - {\mu_y^2 \lambda_y^{-2} } 
	  +
  	1  
  \right)
	  \cos \left( \frac{2\pi}{N} I\right)
  \right]^{-1/2}  \,.
	 \label{Eq:TrrhoNF}
 \end{align}
Using the identity 
\begin{equation}
	\prod_{k=1}^{N-1} 
	\left( \cosh x - \cos \frac{2\pi k}{N}\right) 
	= \frac{1} { 2^{N-1} } \frac{\cosh N x - 1 }{ \cosh x - 1 } \,,
\end{equation}
we obtain 
\begin{equation}
	\notag  {\rm Tr}  \left( \hat\rho \right)^N   = 
	{\rm det} \left[ \frac{1}{2^{2N-1}} \left(  
		{ \mu_y^2 \lambda_y^{-2} } 
		-1 \right)^{N} 
		\left( 
	T_N  
	 \left( \frac{{\mu_y^2 \lambda_y^{-2} } 
 +1   } {
	 \mu_y^2 \lambda_y^{-2} 
 -1 
 }  \right)  
		-1   \right) 
   \right]^{-1/2} ,
	\label{Eg:Beauty}
\end{equation}
where $T_N$ are the  Chebyshev polynomials    
\begin{equation}
T_N\left( x  \right) 
= \cosh \left( N\,  {\rm acosh} \, x\right) \,.
\label{Eq:Tch}
\end{equation}
Thus the $N$-th Renyi entropy is 
\begin{equation}
	S_N = 
	\frac{1}{2 (N-1) }  {\rm tr} 
	\left\{ 
	- (2 N-1) \ln 2 + N \ln   \left(  
		{ \mu_y^2 \lambda_y^{-2} } 
		-1 \right)
	+ \ln 
	\left( T_N \left( \frac{{\mu_y^2 \lambda_y^{-2} } 
 +1   } {
	 {\mu_y^2 \lambda_y^{-2} } 
 -1 
 }  \right)  -1   \right) 
	\right\}. 
	\label{Eq:Sngen}
\end{equation}
Note that this expression gives zero in the limit $  \lambda_y^2 \to  \mu_y^2$; 
this can be easily established based on the property of the leading 
order coefficients of the Chebyshev polynomial of order $N$: $T_N (x\rightarrow \infty) \approx 2^{N-1} x^N
- 2^{N-3} N x^{N-2}$. The second term in this expansion allows to extract the first non-trivial contribution 
to $S_N$ close to the pure state limit  $  \lambda_y^2 \to   \mu_y^2$:
\begin{equation}
	S_N = \frac{N}{4 (N-1) }  {\rm tr}   \left(  
		 \mu_y^2 \lambda_y^{-2}
		-1 \right) 
		+ { \cal O} \left(  \left[  
		 \mu_y^2 \lambda_y^{-2} 
	-1 \right]^2 
  \right)  
		\,.
\end{equation}
In the opposite limit, in a   strongly mixed state,  $ \lambda_y^2 \ll    \mu_y^2$,
we get 
\begin{equation}
	S_N = 
	\frac{1}{2}  {\rm tr} \left[  
	\frac{2 \ln N }{N-1}+
 \ln \left( 	 \mu_y^2 \lambda_y^{-2}   \right) \right]
		+ { \cal O} \left(  \left[  
		 \mu_y^2 \lambda_y^{-2}  \right]^{-1} 
  \right)  
		\,,
		\label{Eq:S_N_mixed}
\end{equation}
which is $N$-independent at  leading order.

The usual  Renyi entropy of the Gaussian density matrix for $N=2$ can be computed directly  to yield  
\begin{equation}
S_R \equiv S_2 =-\ln[{\rm Tr}\hat\rho^2]=\frac12 {\rm tr}\left[\ln \left( \mu_y^2 \lambda_y^{-2} \right) \right]\,.
\end{equation}
The same result can be obtained from  the general Eq.~\eqref{Eq:Sngen} taking $N=2$.
Here, as before,  $\lambda^2$ and $\mu^2$ are  considered as operators on the transverse space. 

It is clear from the above expressions that the increase of $ \mu^2$ and/or decrease of $\lambda^2$ 
increases the Renyi entropy and thus signals an increased mixing of the density matrix.
As we have seen in the previous subsection, this is indeed what happens  in the dilute regime. Using Eq.~(\ref{growth}) we find that, in the dilute regime,
\begin{equation}
\frac{d}{dy}S_R=\gamma\,. 
\end{equation}
That is, the  Renyi entropy grows linearly with  rapidity.

\begin{figure}[h]
\includegraphics[width=0.5\linewidth]{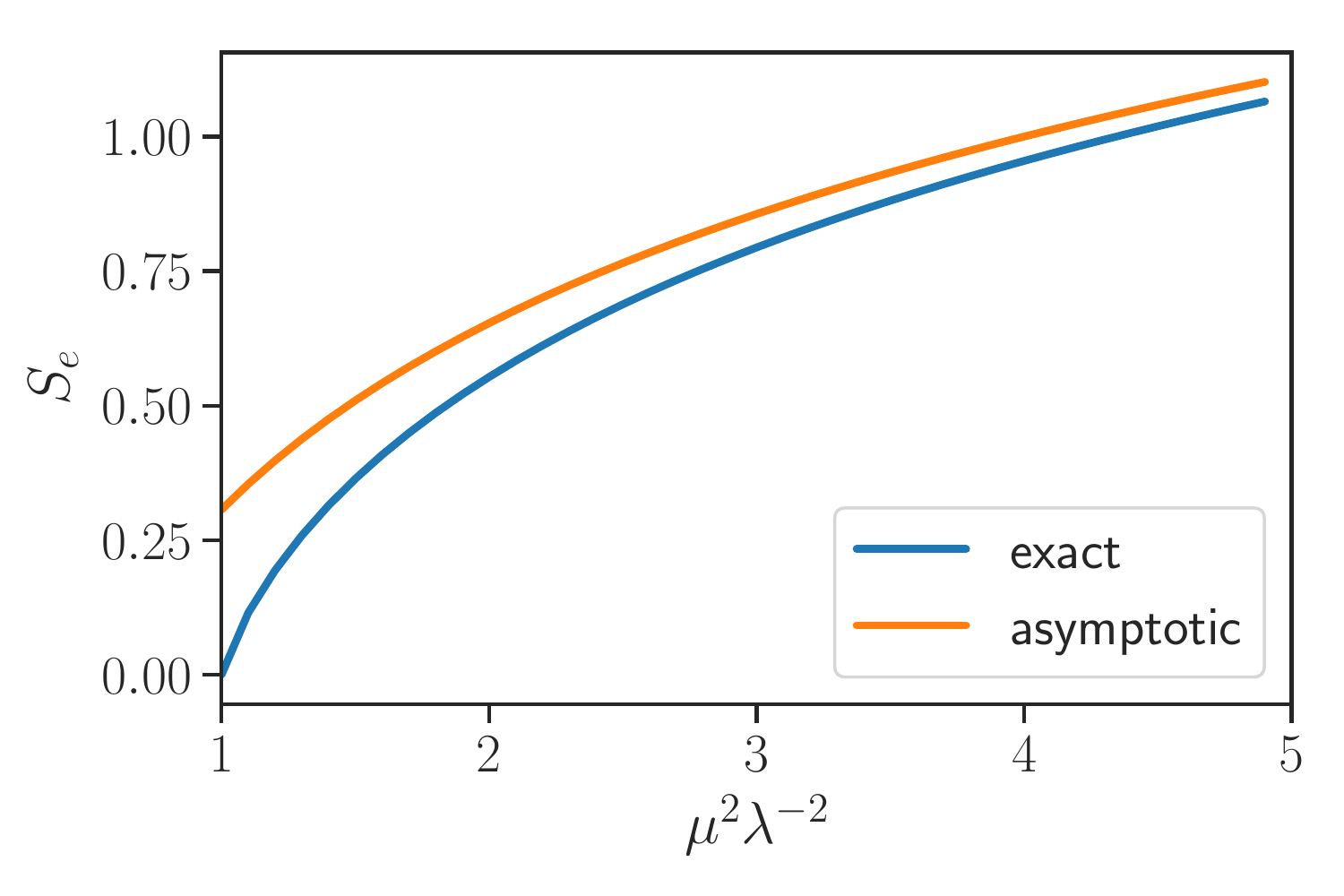}
\caption{ Illustration of the  von Neumann entropy: the function $S_e(x =  \mu_y^2 \lambda_y^{-2} )$, where we treat $\mu_y^2$ and  $\lambda_y$ as scalar numbers. The blue line is the exact result (\ref{eq:entropydilute}) and the orange line its strongly mixed state limit $ \lambda_y^2 \ll   \mu_y^2$.}
\label{fig:S}
\end{figure}

Now consider the von Neumann entropy of the reduced density matrix; it is  defined by 
\begin{equation}
S_e = - {\rm Tr} \left( \hat \rho  \ln \hat \rho  \right)\,.  
\end{equation}
Using the identity $ \ln \hat \rho = \lim_{\epsilon \to 1} \frac{ \hat\rho^{\epsilon-1}  -1} { \epsilon-1 }$ 
we can reduce the evaluation of  the von Neumann entropy to the calculation of $S_1$:
\begin{equation}
	S_e = - \lim_{\epsilon \to 1} {\rm Tr} \left(  \frac{ \rho^\epsilon - \rho} { \epsilon -1}   \right) 
	=    - \lim_{\epsilon \to 1}    \frac{ e^{(1-\epsilon) S_\epsilon} -1} { \epsilon -1} 
\end{equation}
and, assuming that $\lim_{\epsilon\to 1} S_\epsilon$ exists, we obtain 
\begin{equation}
	S_e = S_1 = \frac{1}{2} {\rm tr }\left[\ln \left(\frac{  \mu_y^2 \lambda_y^{-2}      -1}{4}\right)+\sqrt{  \mu_y^2 \lambda_y^{-2}  } {\rm acosh}
   \left(\frac{  \mu_y^2 \lambda_y^{-2}  +1}{  \mu_y^2 \lambda_y^{-2}  -1}\right)\right]  \,. 
   \label{eq:entropydilute}
\end{equation}
This is an exact expression for the  von Neumann entropy of the reduced density matrix in the Gaussian approximation. To demonstrate the behavior of the entropy, in Fig.~\ref{fig:S}, we plotted 
the function $S_e(x = \mu_y^2 \lambda_y^{-2})$, where we treat $\mu_y^2$ and  $\lambda_y$ as scalar numbers.

For a strongly mixed state (i.e. $|\mu^2\lambda^{-2}|\gg 1$), 
the evolution reads:
\begin{equation}\label{bfkles}
	\frac{d S_e}{d y} \approx  
	\frac{1}{2}  {\rm tr}   
	\left[  
		  \mu_y^{-2} 
		\frac{\partial  \mu_y^2}{\partial y} 
		-		 \lambda_y^{-2}
	\frac{\partial  \lambda_y^2}{\partial y} \right]\, .  
\end{equation}
Thus, in the BFKL regime for large enough energy
\begin{equation}\label{bfkle}
	\frac{dS_e}{d y} \approx  
	\gamma \,.  
\end{equation}

In full generality, the evolution of the  von Neumann entropy reads  
\begin{equation}
	\frac{d S_e}{d y} = \frac{1}{4}{\rm tr}\left[ \frac{{\rm acosh}\left(\frac{ \mu_y^2 \lambda_y^{-2}  +1}{ \mu_y^2 \lambda_y^{-2}  -1}\right)}{ \sqrt{ \mu_y^2 \lambda_y^{-2} }}
	\frac{\partial } { \partial y } \left(\mu_y^2 \lambda_y^{-2} \right)\right]\,. 
\end{equation}
Close to the pure state limit this  reduces to   
\begin{equation}
	\frac{d S_e}{d y} \approx 
	- \frac{1}{4}
	 {\rm tr}\left[
	 \ln \left( \frac{\mu_y^2 \lambda_y^{-2}    - 1 }{4} \right)
    	\frac{\partial } { \partial y } \left(\mu_y^2 \lambda_y^{-2}  \right)\right] \,.
\end{equation} 
Therefore, the entropy deviates fast from the pure state regime due to the presence of the logarithmic singularity in the derivative close to the pure  state limit.  This shows that, at least in the Gaussian approximation, a pure state quickly morphs into a mixed state.

%
%
%

\subsection{Approach to saturation and Levin-Tuchin law}
We now wish to study the behavior of the entanglement entropy in the saturated regime. We will again take a reasonable ansatz for the density matrix and will require that it reproduces the evolution of two simple operators. The two operators that we choose are the dipole amplitude in the fundamental representation
\begin{equation}
d(x_{1 \perp},x_{2 \perp})\equiv\frac{1}{N_c}{\rm tr}_c[S^\dagger(x_{1 \perp})S(x_{2 \perp})]
\end{equation}
and the 
correlator
\begin{equation}
P^\dagger(x_{1 \perp},x_{2 \perp})=J^a_R(x_{1 \perp})J^a_R(x_{2 \perp}).
\end{equation}
As explained in Ref. \cite{Altinoluk:2013rua}, the operator $P^\dagger$ (at least in the first approximation) plays the role of the conjugate Pomeron within the Pomeron field theory approximation to  JIMWLK evolution.

Before restricting ourselves to a particular form of $\hat\rho$, let us derive the operator evolution of the two simple operators in question.

\subsubsection{Evolution}
It is straightforward to derive the evolution for operators without making simplifying assumptions about the strength of the gluon fields but instead using the full expression for the jump operator $\hat Q^a(z_\perp)$.
It is in fact obvious that the dipole evolves according to the BK equation~\cite{Balitsky:1995ub,Kovchegov:1999ua}
\begin{equation}
\frac{d}{dy}d(x_{1 \perp},x_{2 \perp})=\frac{\alpha_sN_c}{\pi}\int \frac{d^2z_\perp}{2\pi} \frac{(x_{1 \perp}-z_\perp) \cdot(x_{2 \perp}-z_\perp) }{(x_{1 \perp}-z_\perp) ^2(x_{2 \perp}-z_\perp) ^2}\left[d(x_{1 \perp},z_\perp)d(z_\perp, x_{2 \perp})-d(x_{1 \perp},x_{2 \perp})\right]\,. 
\end{equation}
The reason is that the operator $d$ depends only on the eikonal matrices $S$, and thus its average and evolution is governed entirely by the diagonal elements of $\hat\rho$ in the $S$-basis. Since these elements evolve according to the original JIMWLK equation, so does the operator $d$.

The  explicit calculation for  $P^\dagger$ yields
\begin{eqnarray}
&&\int d^2z_\perp[[J^b_R(x_{2 \perp})J^b_R(x_{1 \perp}),\hat Q^a_i(z_\perp)],\hat Q^a_i(z_\perp)]=\frac{g^2}{(2\pi)^2}\\
&&\Bigg[\int d^2x_\perp\frac{(x_\perp- x_{1 \perp})\cdot(x_{2 \perp}-x_{1 \perp})}{(x_\perp- x_{1 \perp})^2(x_{2 \perp}-x_{1 \perp})^2}J_L^d(x_{2 \perp})[S(x_{2 \perp})T^bS^T(x_{1 \perp})]^{da}J^c_R(x_\perp)[T^bS^T(x_{1 \perp})]^{ca}\nonumber\\
&&+\delta^{(2)}(x_{1 \perp}-x_{2 \perp})\int d^2x_\perp d^2y_\perp\frac{(x_\perp- x_{1 \perp})\cdot(y-x_{2 \perp})}{(x_\perp- x_{1 \perp})^2(y-x_{2 \perp})^2}J_L^d(y_\perp)[S(y_\perp)T^bS^T(x_{2 \perp})]^{da}J^c_R(x_\perp)[T^bS^T(x_{2 \perp})]^{ca}\nonumber\\
&&-\int d^2z_\perp\frac{(x_{1 \perp}-z_\perp) \cdot(x_{2 \perp}-z_\perp) }{(x_{1 \perp}-z_\perp) ^2(x_{2 \perp}-z_\perp) ^2}J_L^d(x_{2 \perp})[S(x_{2 \perp})T^bS^T(z_\perp)]^{da}J^c_R(x_{1 \perp})[T^bS^T(z_\perp)]^{ca}\nonumber\\
&&-\int d^2y_\perp\frac{(x_{1 \perp}-x_{2 \perp})\cdot(y-x_{2 \perp})}{(x_{1 \perp}-x_{2 \perp})^2(y-x_{2 \perp})^2}J_L^d(y_\perp)[S(y_\perp)T^bS^T(x_{2 \perp})]^{da}J^c_R(x_{1 \perp})[T^bS^T(x_{2 \perp})]^{ca}\nonumber\\
&&+\int d^2x_\perp\frac{1}{(x_\perp- x_{1 \perp})^2}J_R^b(x_{2 \perp})J_L^d(x_\perp)[S(x_\perp)T^cS^T(x_{1 \perp})]^{da}[T^bS^T(x_{1 \perp})]^{ca}\nonumber\\
&&-\int d^2z_\perp\frac{1}{(x_{1 \perp}-z_\perp) ^2}J_R^b(x_{2 \perp})J_L^d(x_{1 \perp})[S(x_\perp)T^cS^T(z_\perp)]^{da}[T^bS^T(z_\perp)]^{ca}\Bigg]\nonumber\\
&&+(x_{2 \perp}\leftrightarrow x_{1 \perp}).\nonumber
\end{eqnarray}
To simplify this, we use $S^TS=1$, $(T^a)^T=-T^a$ , $J_LS=J_R$ and $T^aT^a=N_c$, getting
\begin{eqnarray}
&&\frac{d}{dy}P^\dagger(x_{1 \perp},x_{2 \perp})=\int \frac{d^2z_\perp}{2\pi} \left[[P^\dagger(x_{1 \perp},x_{2 \perp}),\hat Q^a_i(z_\perp)],\hat Q^a_i(z_\perp)\right]\\&&=-\frac{g^2N_c}{(2\pi)^3}\Bigg[\delta(x_{1 \perp}-x_{2 \perp})\int d^2x_\perp d^2y_\perp\frac{(x_\perp- x_{1 \perp})\cdot(y-x_{2 \perp})}{(x_\perp- x_{1 \perp})^2(y-x_{2 \perp})^2}P^\dagger(x_\perp,y_\perp)\nonumber\\
&&+\int d^2x_\perp\left[\frac{(x_\perp- x_{1 \perp})\cdot(x_{2 \perp}-x_{1 \perp})}{(x_\perp- x_{1 \perp})^2(x_{2 \perp}-x_{1 \perp})^2}- \frac{1}{(x_\perp- x_{1 \perp})^2}\right] P^\dagger(x,x_{2 \perp})\nonumber\\
&&-\int d^2z_\perp\left[\frac{(x_{1 \perp}-z_\perp) \cdot(x_{2 \perp}-z_\perp) }{(x_{1 \perp}-z_\perp) ^2(x_{2 \perp}-z_\perp) ^2}-\frac{1}{(x_{1 \perp}-z_\perp) ^2}\right]P^\dagger(x_{1 \perp},x_{2 \perp}) \nonumber\\
&&-\int d^2x_\perp\frac{(x_{1 \perp}-x_{2 \perp})\cdot(x_\perp- x_{2 \perp})}{(x_{1 \perp}-x_{2 \perp})^2(x_\perp- x_{2 \perp})^2}P^\dagger(x_{1 \perp},x_\perp)\Bigg]\nonumber\\
&&+(x_{1 \perp}\leftrightarrow x_{2 \perp})\nonumber \,. 
\end{eqnarray}

Interestingly we find that even {\it in the saturated regime} the charge density correlator evolves according to the BFKL equation.
This is perhaps not completely surprising for the following reason. As discussed in the literature, e.g. Ref.~\cite{Kovner:2005en}, high energy evolution has a self dual structure. As a result of this dense-dilute duality, the operators that depend on the eikonal matrix $S$ probe the structure of the target state, while those that depend on the charge operators $J$ effectively probe the structure of the projectile state. The JIMWLK evolution describes a situation where the target is dense, but the projectile is dilute. Thus the rapidity evolution of the charge correlators reflects the rapidity evolution of charge densities in the dilute projectile, which have to evolve according to the BFKL equation. This indeed is what we find. Note, however, that even though this result is natural, it by no means trivial, as we were only able to obtain it explicitly by using the density matrix formulation of  high energy evolution, as the knowledge of the diagonal matrix elements of $\hat\rho$  alone is not sufficient to calculate the evolution of any function of $J$'s.

\subsection{The ansatz for $\hat\rho$ in the saturated regime and the operator averages}

To study the density matrix close to the saturated regime we will  take a natural generalization of the Gaussian ansatz ($S$ and $\bar S$ below are matrices in the fundamental representation),
\begin{align}
\hat\rho(S,\bar S)={\cal N}\exp \Bigg\{&-
{\rm tr}_{\rm c}
\int d^2x_\perp d^2y_\perp 
\Big[\frac{\bar\mu_y^{-2}(x_\perp,y_\perp)}{4}[S^\dagger(x_\perp)+\bar S^\dagger(x_\perp)][S(y_\perp)+\bar S(y_\perp)] \notag \\ &+ \bar\lambda_y^{-2}(x_\perp,y_\perp)[S^\dagger(x_\perp)-\bar S^\dagger(x_\perp)][S(y_\perp)-\bar S(y_\perp)]\Big]\Bigg\},
\end{align}
and repeat the procedure that we performed in  the dilute regime.  Note that we already made  the unitary transformation to rotate out the function $A$, as in the dilute case. 
Thus, similarly to the dilute regime, we have to consider only two operators in order to derive  the evolution of parameters $\bar\lambda$ and $\bar\mu$.  

For simplicity, we will adopt the following natural assumptions   $\bar\mu_y^2(x_\perp,y_\perp)=\bar\mu_y^2(y_\perp,x_\perp)$ and $\bar\lambda_y^2(x_\perp,y_\perp)=\bar\lambda_y^2(y_\perp,x_\perp)$. This prevents the appearance of the odderon which is not critical for our consideration relevant for high energy. In general, one can lift this assumption and repeat the derivation; for the purpose of this paper it is not necessary. 

Our goal is now to calculate the averages of $d$ and $P^\dagger$ in this density matrix.
This is not an easy task and we do not know how to perform this calculation in full generality. 
 However, since we are interested in the behavior close to the saturation limit, we 
can invoke the factorized approximation used in Ref.~\cite{Kovner:2017ssr,Altinoluk:2018ogz,Kovner:2018azs}.  
This amounts to forgetting about the complicated group measure while integrating over $S$, and using the standard measure on complex numbers $C$  for each matrix element of $S$. 
In this approximation, the averages of products of $S$ matrices factorize into products of color singlet pairs, see Ref.~\cite{Kovner:2017ssr,Altinoluk:2018ogz,Kovner:2018azs} for details.
This approximation is justified in particular when one is interested in leading powers of the area of the projectile as explained in the above references. We will not further justify this approximation here, but will instead hope that it gives qualitatively correct answers to the questions that we are asking.

Our normalization of $\bar\mu^2$ is such that
\begin{equation}
d(x_{1 \perp},x_{2 \perp})\equiv \left\langle \frac{1}{N_c}{\rm tr}_c\, [S^\dagger(x_{1 \perp})S(x_{2 \perp})]\right \rangle=N_c\bar\mu^2(x_{1 \perp},x_{2 \perp}) \,. 
\end{equation}
This means that the natural magnitude is $\bar\mu^2\sim 1/N_c$.

To calculate $\langle P^\dagger\rangle$ we will use the identities 
\begin{equation}
J^a_R(x_\perp)S(y_\perp)=\delta^{(2)}(x_\perp-y_\perp)S(x_\perp)T^a \ , \,\,\,\, ~~ J^a_R(x_\perp)S^\dagger(y_\perp)=-\delta^{(2)}(x_\perp-y_\perp)T^aS^\dagger(x_\perp),
\end{equation}
which can be trivially proven based on the definition of $J^a_R(x_\perp)$, see Eq.~\eqref{Eq:JR}. Using these identities we obtain 
\begin{align}
&J^a_R(x_{1 \perp})\hat\rho(S,\bar S)=-\Big[\int d^2x_\perp \frac{\bar\mu_y^{-2}(x,x_{1 \perp})}{4}{\rm tr}_c[S^\dagger(x_\perp)+\bar S^\dagger(x_\perp)][S(x_{1 \perp})T^a]\nonumber \\
&-\frac{\bar\mu^{-2}(x_{1 \perp},x_\perp)}{4}[T^aS^\dagger(x_{1 \perp})][S(x_\perp)+\bar S(x_\perp)]\nonumber\\
&+\bar\lambda^{-2}(x,x_{1 \perp}){\rm tr}_c[S^\dagger(x_\perp)-\bar S^\dagger(x_\perp)][S(x_{1 \perp})T^a]-\bar\lambda^{-2}(x_{1 \perp},x_\perp)[T^aS^\dagger(x_{1 \perp})][S(x_\perp)-\bar S(x_\perp)]\Big]\hat\rho(S,\bar S).
\end{align}
Before acting with the second operator $J^a_R$, we note that to calculate the average we will have to set $\bar S=S$ after the differentiation. We will therefore only keep terms that do not vanish for $\bar S=S$:
\begin{eqnarray}
&&J^a_R(x_{2 \perp})J^a_R(x_{1 \perp})\hat\rho(S,\bar S)|_{\bar S=S}=\hat\rho(S,S)\Big\{ C_F\left(\frac{\bar\mu^{-2}(x_{2 \perp},x_{1 \perp})}{4}+\bar\lambda_y^{-2}(x_{2 \perp},x_{1 \perp})\right)\nonumber \\
&& \hskip 3cm \times {\rm tr}_c[S^\dagger(x_{2 \perp})S(x_{1 \perp})+S^\dagger (x_{1 \perp})S(x_{2 \perp})]\nonumber\\
&&\hskip 2cm -C_F\,\delta^{(2)}(x_{1 \perp}-x_{2 \perp})\int d^2x_\perp \frac{\bar\mu^{-2}(x,x_{1 \perp})}{2}{\rm tr}_c[S^\dagger(x_\perp)S(x_{1 \perp})+S^\dagger (x_{1 \perp})S(x_\perp)]\Big\}\\
&&\hskip 1cm +\int d^2x_\perp d^2y_\perp \frac{\bar\mu^{-2}(x,x_{1 \perp})}{2}\frac{\bar\mu^{-2}(y_\perp, x_{2 \perp})}{2}\left[{\rm tr}_c[S^\dagger(x_\perp)S(x_{1 \perp})T^a-S^\dagger(x_{1 \perp})S(x_\perp)T^a\right]\nonumber \\
&& \hskip 3cm \times\left[{\rm tr}_c[S^\dagger(y_\perp)S(x_{2 \perp})T^a-S^\dagger(x_{2 \perp})S(y_\perp)T^a\right]\nonumber\\
&=&\hat\rho(S,S)\Big\{C_F\left(\frac{\bar\mu^{-2}(x_{2 \perp},x_{1 \perp})}{4}+\bar\lambda^{-2}(x_{2 \perp},x_{1 \perp})\right){\rm tr}_c[S^\dagger(x_{2 \perp})S(x_{1 \perp})+S^\dagger (x_{1 \perp})S(x_{2 \perp})]\nonumber\\
&-&C_F\delta(x_{1 \perp}-x_{2 \perp})\int d^2x_\perp \frac{\bar\mu^{-2}(x,x_{1 \perp})}{2}{\rm tr}_c[S^\dagger(x_\perp)S(x_{1 \perp})+S^\dagger (x_{1 \perp})S(x_\perp)]\Big\}\nonumber \\
&+&\int d^2x_\perp d^2y_\perp \frac{\bar\mu^{-2}(x,x_{1 \perp})}{2}\frac{\bar\mu^{-2}(y_\perp, x_{2 \perp})}{2}\Big\{{\rm tr}_c[(S^\dagger(x_\perp)S(x_{1 \perp})-S^\dagger(x_{1 \perp})S(x_\perp))(S^\dagger(y_\perp)S(x_{2 \perp})-S^\dagger(x_{2 \perp})S(y_\perp))]\nonumber\\
&&\ \ \ \ \ \ \ \ \ \ \ \ \ \ \ \ \ \ \ \ \ \ \ \ \ \ \ \ \ \ \ \ \ \  \ \ \ -\frac{1}{N_c}{\rm tr}_c[S^\dagger(x_\perp)S(x_{1 \perp})-S^\dagger(x_{1 \perp})S(x_\perp)]{\rm tr}_c[S^\dagger(y_\perp)S(x_{2 \perp})-S^\dagger(x_{2 \perp})S(y_\perp)]\Big\}\nonumber \, .
\end{eqnarray}

Now recall that the factorization rules dictate 
\begin{align}
	\langle{\rm tr}_c[S^\dagger(x_{1 \perp})S(x_{2 \perp})S^\dagger(x_3)S(x_4)]\rangle  \approx \frac{1}{N_c}\Bigg\{& {\rm tr}_c\langle[S^\dagger(x_{1 \perp})S(x_{2 \perp})]\rangle{\rm tr}_c[\langle S^\dagger(x_3)S(x_4)\rangle] \notag \\ &+{\rm tr}_c[\langle S^\dagger(x_{1 \perp})S(x_4)\rangle]{\rm tr}_c[\langle S^\dagger(x_3)S(x_{2 \perp})]\rangle\Bigg\}\, .
\end{align}

For simplicity we calculate the averages to leading order in $1/N_c$. We then find
\begin{equation}
\langle J^a_R(x_{2 \perp})J^a_R(x_{1 \perp})\rangle=2N_c^3\bar\lambda_y^{-2}(x_{1 \perp},x_{2 \perp})\bar\mu_y^2(x_{1 \perp},x_{2 \perp}),
\end{equation}
which suggests that $\bar\lambda_y^{-2}$ is of order 1.

In order to restore the natural $N_c$ power counting, we rescale $\bar\mu^2\rightarrow \frac{\bar\mu^2}{N_c}$ and obtain 
\begin{equation}
d(x_{1 \perp},x_{2 \perp})=\bar\mu_y^2(x_{1 \perp},x_{2 \perp}),\ \ \ \ \ \ \ \ \frac{1}{N_c^2}\langle J^a_R(x_{2 \perp})J^a_R(x_{1 \perp})\rangle=2\bar\lambda^{-2}(x_{1 \perp},x_{2 \perp})\bar\mu_y^2(x_{1 \perp},x_{2 \perp}) \,. 
\end{equation}

 Recall that in the saturation regime the behavior of the dipole is governed by the Levin-Tuchin (LT) formula~\cite{Levin:1999mw}  
\begin{equation}
d(x_{1 \perp},x_{2 \perp})=\exp\{-\xi\ln^2[(x_{1 \perp}-x_{2 \perp})^2Q_s^2]\}\, , 
\end{equation} 
where $\xi$ is a constant of order unity, and $Q_s$ is the saturation momentum.
We thus conclude that in this regime
\begin{equation}\label{mu}
\bar\mu_y^2(x_{1 \perp},x_{2 \perp})=\exp\{-\xi\ln^2[(x_{1 \perp}-x_{2 \perp})^2Q_s^2]\} \,. 
\end{equation}
Given that the color density correlator satisfies the BFKL equation, we find
\begin{equation}
\bar\lambda_y^{-2}(x_{1 \perp},x_{2 \perp})\bar\mu_y^2(x_{1 \perp},x_{2 \perp})\approx \bar\lambda_0\exp (\gamma y)
\end{equation}
or 
\begin{equation}\label{lambda}
\bar\lambda_y^{-2}(x_{1 \perp},x_{2 \perp})=\bar\lambda_0\exp\left\{\gamma y+\xi\ln^2[(x_{1 \perp}-x_{2 \perp})^2Q_s^2(y_\perp)] \right\} \,, 
\end{equation}
where $\bar\lambda_0$ is determined by the initial condition.
The dependence of $Q_s^2$ on $y$ is well known, with leading exponential behavior being
\begin{equation}
Q_s^2(y_\perp)=Q_s^2e^{\beta y},\ \ \ \ \ \beta=\frac{\alpha_sN_c}{\pi}\frac{\chi(\gamma_c)}{\gamma_c}\,,
\end{equation}
where $\chi$ is the BFKL kernel, $\chi(\gamma)=2\psi(1)-\psi(\gamma)-\psi(1-\gamma)$, and $\gamma_c$ is the solution of equation
$\chi(\gamma_c)=\gamma_c\chi'(\gamma_c)$, which numerically is $\gamma_c\approx .628$.
Numerically $\beta\approx 4.88\frac{\alpha_sN_c}{\pi}$~\cite{Iancu:2002aq}. 

Note that the density matrix is normalizable within our approximation  only as long as $|\bar\lambda^{-2}|>\frac{N_c}{4}|\bar\mu^{-2}|$. 
Thus the calculation can only be valid for  large rapidities, i.e.
\begin{equation}\label{limit}
e^{\gamma y}>\frac{N_c}{\bar\lambda_0} \,. 
\end{equation}
This is quite reasonable. Recall that the saturation regime sets in parametrically when
\begin{equation}
e^{\gamma y}\alpha_s\sim 1.
\end{equation}
Since at large $N_c$ we have $\alpha_s\sim 1/N_c$, parametrically this is the same as Eq.~(\ref{limit}) if the initial condition $\bar\lambda_0$ is of the order of the 't Hooft coupling, $\bar\lambda_0\sim \alpha_sN_c$. 

\subsection{Entropy in the saturated regime}
The next natural question is how does entropy evolve in the LT regime. We can in fact adopt the results of the previous section to calculate entropy. Our approximation of calculating the functional integral corresponds simply to treating the matrix elements of $S$ as independent degrees of freedom. The density matrix therefore behaves as a Gaussian in these degrees of freedom and the entropy is simply the entropy of a Gaussian density matrix. We therefore can directly write
\begin{equation}
	S_e = \frac{1}{2} {\rm Tr }\left[\ln \left(\frac{   4 \bar\mu_y^2 \bar\lambda_y^{-2} / N_c       -1}{4}\right)+\sqrt{  4 \bar\mu_y^2 \bar\lambda_y^{-2} / N_c  }\,  {\rm acosh}
	\left(\frac{4 \bar\mu_y^2 \bar\lambda_y^{-2}/ N_c  +1}{4 \bar\mu_y^2\bar\lambda_y^{-2}/N_c   -1}\right)\right] \,.  
\end{equation}
Here the matrices $\bar\mu^2$ and $\bar\lambda^2$ are Hermitian $N_c^2$ by $N_c^2$ matrices.

To estimate this we need to calculate the operator
\begin{equation}
M(y_\perp, z_\perp)=\int d^2x_\perp \bar\mu^2(y_\perp, x_\perp)\bar\lambda^{-2}(x_\perp,z_\perp)=e^{\gamma y}\int d^2x_\perp e^{-\beta[\ln^2(x_\perp-y_\perp)^2-\ln^2(x_\perp-z_\perp)^2]} \,. 
\end{equation}
For very large target area ${\cal A}$ the integral is obviously dominated by the values of $x_\perp$ very far from $y_\perp$ and $z_\perp$, and we obtain
\begin{equation}
M(y_\perp, z_\perp)\approx {\cal A}e^{\gamma y}.
\end{equation}
We can now recast Eq.~(\ref{bfkle}) in the form
\begin{equation}  \label{enev}
\frac{d S_e}{d y}\approx \frac{1}{2}\gamma \, .
\end{equation}
Thus, interestingly the entropy grows slower in the saturated regime.
This is natural since due to saturation effects the emission of soft gluons is suppressed and, thus, one expects the rate of decoherence of the density matrix to slow down.

\section{Wigner functional}

Let us now return to our original motivation for introducing the density matrix: can we get information on the distribution of currents in the hadronic state at high energy and, more interestingly,  on correlations between currents and color charge densities? This type of question is particularly pertinent as we are interested in rare configurations in the wave function, e.g. such that produce higher than average  multiplicity final states in p-p collisions. Such configurations are quite likely to also harbor large currents and therefore momentum distributions that significantly differ from the average.

A similar question in a single particle quantum mechanics is answered, at least partially, by the Wigner function, which can be approximately interpreted 
as giving the joint probability for the distribution of position and momentum of the particle,
\beq
{\cal W}(x,p)=\int dy e^{iyp} \left \langle x+\frac{y}{2}\Big|\hat\rho\Big|x-\frac{y}{2} \right \rangle \,  . 
\eeq
The momentum is proportional to the velocity of the particle, and thus the Wigner function carries information not just about the distribution of position but also about its time derivative. 
This joint distribution is a very interesting quantity since it can, among other things, tell us how fast the particle escapes from a given point in space.

One can define an analog of the Wigner function for a field theory. Formally let us define the Wigner functional as
\beq\label{wifi}
{\cal W}[j, \Phi]=\int Dj' \exp  \left[ i \int d^2x_\perp \Phi(x_\perp) j' (x_\perp) \right]
\rho\left[j+\frac{j'}{2},j-\frac{j'}{2}\right] \,. 
\eeq
The high energy evolution of this functional can be readily derived from the evolution of $\hat\rho$. We will not pursue this trivial derivation here but instead concentrate on a possible 
phenomenological application of the functional. 

Does this functional give us any information about the distribution of color current densities, as opposed to just the distribution of color charge densities? 
The color current density operator is not directly present in the effective description furnished by $\hat\rho$ or ${\cal W}$ in Eq. (\ref{wifi}). However, as it is usually the case with effective theories, certain fundamental operators can be related to objects appearing in the effective description. The only object independent of $j$ that appears in Eq. (\ref{wifi}) is the phase $\Phi$.
Thus we need to understand if $\Phi$ is related to the current density. 

Eq.~(\ref{wifi}) defines $\Phi$ as the canonical conjugate of $j$, i.e.
\beq\label{phi}
\Phi=-i\frac{\delta}{\delta j} \,. 
\eeq
 
 Let us calculate the commutation relations between the color current and the color charge density in the fundamental description. Recall that on the ``microscopic'' level we have the color current density~\footnote{We omit for simplicity the longitudinal coordinate label of the creation and annihilation operator. It can be checked trivially that reinstating it does not change our results.}
\beq j_i^a(x_\perp)=\frac{1}{2}f^{abc}[a^{\dagger b}_l(x_\perp)\partial_ia^c_l(x_\perp)-\partial_ia^{\dagger b}_l(x_\perp)a^c_l(x_\perp)] \, . 
\eeq

Commuting this with the color charge density
\beq 
j ^a(x_\perp)=i f^{abc} a^{\dagger b}_l(x_\perp)  a^c_l(x_\perp),
\eeq
we get 
\beq\label{algebra}
[j^a(y_\perp),j_i^b(x_\perp)]=if^{abc}j^c_i(x_\perp)\delta^{(2)}(x_\perp-y_\perp)+if^{abc}j^c(x_\perp)\partial^y_i\delta^{(2)}(x_\perp-y_\perp) \, , 
\eeq
where we used the canonical commutation relations 
\begin{equation}
	[a_i^a(x_\perp), a_j^b(y_\perp) ] = \delta_{ij} \delta^{ab} \delta^{(2)} (x_\perp-y_\perp)	
\end{equation}
and the Jacobi identity for the structure constants.

Given this commutation relation and Eq.~(\ref{phi}), we can construct operators in the effective theory which satisfy the same algebra. In particular, to reproduce Eq.~(\ref{algebra}) we can adopt the following representation for the current density in the ``effective'' description:
\beq 
 j_i^a(x_\perp)=f^{abc}j^b(x_\perp)\partial_i\Phi^c(x_\perp) \,. 
 \eeq
 
 In other words, indeed if we have a joint probability distribution of $j$ and $\Phi$, we also know the joint probability distribution of $j$ and $j_i$. 
 
 The color charge satisfies a covariant conservation equation. Disregarding for the moment the word ``covariant", this amounts to 
 \beq
 p^+\partial^-j=\partial_i  j_i \,. 
 \eeq
 This equation, in principle, tells us how fast the charge density ``runs away'' from any given configuration. This runaway speed is of course proportional to $1/p^+$, which is small. But this overall scaling is simply the consequence of Lorentz time dilation in the boosted frame. 

As a simple example of possible utility of the Wigner functional let us do the following simple exercise. We take a  Gaussian ansatz for the density matrix Eq.~(\ref{gauss0}), and calculate the correlation between the color charge and the color current densities. We take $A=0$ for now, and obtain the Wigner functional: 
\begin{equation}
	{\cal W}_G [j, \Phi]  = {\cal N} 
	\exp\left[  - \int d^2x_\perp d^2y_\perp 4 \mu^{-2}(x_\perp,y_\perp) j(x_\perp)  j(y_\perp)  - \frac{1}{4} \int d^2x_\perp d^2y_\perp  \lambda^{2}(x_\perp,y_\perp) \Phi(x_\perp)  \Phi(y_\perp) \right].
	\label{Eq:WG}
\end{equation}
The normalization constant  ${\cal N}$
 can be obtained from the condition 
\begin{equation}
	\int Dj D\Phi\,  {\cal W}_G[j,\Phi] = 1.   
\end{equation}
As one could have expected, the Wigner functional is Gaussian for the Gaussian density matrix. Using this result we can study different  correlators between  color currents and color densities. 
Below we will consider two examples. 
First, we start from correlators of the currents at two different positions, 
\begin{equation}
 	\langle j^a_i (x_\perp) j^b_j (y_\perp)  \rangle   \equiv \int Dj D\Phi\,  {\cal W}_G[j,\Phi]   j^a_i (x_\perp) j^b_j (y_\perp)
=	  N_c \delta_{a,b}\,  \mu^2(x_\perp, y_\perp) \partial^{x_\perp}_i \partial^{y_\perp}_j \lambda^{-2} (x_\perp,y_\perp)\,. 
\end{equation}
This correlator in a way is a proxy for two gluon azimuthal anisotropy harmonics $v_{2n}$ in the CGC wave function. Since it is  proportional to $\lambda^{-2}$,  this demonstrates the importance of the off-diagonal 
components of the full density matrix in relation to the momentum distribution of particles.
 
Another illuminating example potentially pertinent  to phenomenology is the correlator 
\begin{align}\label{crosscorr}
&	\langle j^a (x_\perp) j^a(y_\perp) j^b_i (z_\perp) j^b_i (w_\perp) \rangle -   	\langle j^a (x_\perp) j^a(y_\perp) \rangle \langle  j^b_i (z_\perp) j^b_i (w_\perp) \rangle \notag \\ &
	= \frac{N_c(N_c^2-1)}{4}  \left[ \mu^2(x_\perp,z_\perp) \mu^2(y_\perp,w_\perp) +   \mu^2(x_\perp,w_\perp) \mu^2(y_\perp,z_\perp)    \right]  
	\partial_i^{z_\perp} \partial_i^{w_\perp} \lambda^{-2}(z_\perp,w_\perp) \,. 
\end{align}
Eq.~(\ref{crosscorr}) demonstrates the presence of a nontrivial correlation between a proxy for the gluon multiplicity and the azimuthal anisotropy even in a simple density matrix. Again, to establish this correlation 
the computation of the full density matrix and not just its diagonal part was required. 

One feature of Eq.~(\ref{crosscorr}) is particularly interesting. Note that the correlated (connected) part of the correlator has the same energy dependence as the disconnected piece. Thus this type of correlation, if present in the wave function at initial energy, is not washed away by energy evolution.

Another interesting point is the role of the parameter $A$. If we reinstate it in the general Gaussian ansatz, we obtain for the Wigner functional
\begin{equation}
	{\cal W}_G [j, \Phi]  = {\cal N} 
	\exp\left[  - \int d^2x_\perp d^2y_\perp  4\mu^{-2}(x_\perp,y_\perp) j(x_\perp)  j(y_\perp)  - \frac{1}{4} \int d^2x_\perp d^2y_\perp  \lambda^{2}(x_\perp,y_\perp) \Phi'(x_\perp)  \Phi'(y_\perp) \right],  
	\label{Eq:WG1}
\end{equation}
where we introduced 
\begin{equation}
	\Phi'(x_\perp) = \Phi(x_\perp) + 2 \int d^2 y_\perp j(y_\perp) A(y_\perp,x_\perp)  .
\end{equation}
Although the presence of $A$ has no effect on the correlators involving only $j$, it does affect $j_i$. This is entirely analogous to how a coordinate dependent phase of a wave function does not affect the probability distribution of coordinates, but has a strong effect on the distribution of momenta (velocities) of a particle. Whether this effect is significant at high energy is an interesting question worth exploring.

The upshot of this short discussion is that  the Wigner functional  does have a potential of being a useful tool in understanding the dynamical structure of the hadronic wave function.  
The quantitative study of  the properties and the evolution of the Wigner functional deserves a serious effort, and is left for future work.

\section{Discussion}
In this paper we have introduced the notion of the CGC density matrix $\hat\rho$. This is the reduced density matrix in the CGC effective theory obtained by tracing over all the degrees of freedom
in the QCD Hilbert space except the rapidity integrated color charge density. We stress again that this is not the same density matrix as considered in Refs.~\cite{Kovner:2015hga,Kovner:2018rbf}, where the valence degrees of freedom were integrated out to obtain the reduced density matrix on the soft gluon Hilbert space.

We have derived the evolution equation for the density matrix  $\hat\rho$ and have shown that it is of the Kossakowsky-Lindblad form with the jump operator being equal to the single soft gluon production amplitude. This is intuitively quite agreeable, since in general the meaning of the jump operator is to introduce a jump to a different quantum state of the ``environment'', which in our case contains soft gluon degrees of freedom.

The Kossakowski-Lindblad form is the most general form of {\it Markovian} evolution allowed by a probabilistic interpretation of the density matrix, i.e. overall normalization and positivity of all eigenvalues. This suggests that the general form of the evolution equation should persist beyond leading order. It is thus possible that one can simplify the derivation of the NLO JIMWLK~\cite{Kovner:2013ona,Balitsky:2013fea,Kovner:2014lca,Lublinsky:2016meo} by directly calculating corrections to the jump operator, rather than to the JIMWLK Hamiltonian, which is a more complicated object. One may hope that the same framework can also accommodate  improved leading order JIMWLK versions which resum large transverse logarithms. Physically one expects that since the evolution in energy is aligned with the evolution in the frequency of produced gluons, the typical time scale of the evolution will remain always larger than the time scale of the soft gluon fluctuations. If this is the case, the Markovian nature of the evolution should be preserved beyond the leading order. Although physically reasonable, a better understanding of possible sources for non-Markovian effects in the evolution is necessary.

The Kossakowski-Lindblad evolution is known to lead to increasing entanglement entropy with the evolution ``time''. We have indeed calculated the evolution of entanglement entropy in a Gaussian approximation, both in the dilute regime and close to saturation. We found that in both cases the entanglement entropy increases linearly with rapidity. In the dilute regime the rate of increase coincides with the leading BFKL eigenvalue, while in the dense (Levin-Tuchin) regime it is half of that value. The slower growth of entropy in the saturated regime is likely caused by the suppressed emission probability of soft gluons close to saturation.

The linear growth of entropy with rapidity is a rather interesting result.  One may naively expect that the entropy associated with $\hat\rho$ is proportional to the total gluon number $n_y$ -- at least as long as $n_y$ is not too large. However this is not the case. The total number of gluons in the dilute regime grows with rapidity exponentially, while the entropy Eq.~(\ref{bfkle}) only grows linearly and thus much more slowly. The same type of behavior persists in the dense regime.

A similar behavior of the entanglement entropy of the proton in the context of DIS  was proposed in Ref.~\cite{Kharzeev:2017qzs}. The picture of Ref.~\cite{Kharzeev:2017qzs} is very simple. It assumes that all partonic states in the proton wave function at high energy completely decohere from each other and all accessible states become equally probable. Thus the density matrix becomes proportional to the unit matrix on the subspace of the Hilbert space which is ``populated" at a given energy. The dimension of this subspace is proportional to the mean number of gluons in the wave function, which grows with the BFKL exponential, $d\propto e^{\gamma y}$.   The normalization of $\hat\rho$ means that it has $e^{\gamma y}$ equal eigenvalues, each one approximately $\rho_i\propto d^{-1}$. For such a density matrix we know that the entropy
$S_e\approx -\ln \rho_i=\gamma y$. The growth of this entropy with energy is slow because $\hat\rho$ is already maximally mixed on the subspace of dimension $d$, and the growth is only due to the increase of this dimension with energy. 

Although we do not know how closely the density matrix introduced in the present paper is related to the object considered in Ref.~\cite{Kharzeev:2017qzs}, it is instructive to examine our formulae and understand whether the behavior we find indeed conforms with this simple argument.

Consider for simplicity the density matrix Eq.~(\ref{gauss1}).  Although above we have used this Gaussian ansatz only in the dilute regime, the following qualitative discussion should apply to both regimes. At very high energy, i.e. close to saturation, $\lambda^{-2}_y$ is very large and the density matrix Eq.~(\ref{gauss}) indeed becomes very close diagonal  $\rho(\alpha,\alpha')\propto \delta(\alpha-\alpha')$. Let us for the moment assume that we indeed can neglect the nondiagonal matrix elements of $\hat \rho$. 


Then since $\mu^2_y$ is also very large, for values of the field $\alpha$ such that $\alpha^2<\mu^2_y$ the matrix elements of $\hat\rho$ do not depend on $\alpha$. Thus, in the high energy regime $\hat \rho$ in effect is proportional to a unit matrix of dimension $d\propto |\alpha_{\rm max}|\propto e^{\frac{\gamma}{2} y}$. The entropy associated with such a density matrix should be given by $\ln d=\frac{\gamma}{2}y$. Although qualitatively correct, we are missing here a factor of $1/2$ relative to our result Eq.~(\ref{bfkle}). A closer look at our derivation indeed reveals the origin of the missing factor $1/2$.
 As is obvious from Eq.~(\ref{bfkles}), only half of the entropy growth comes from the growth of $\mu^2$ and therefore of the dimension $d$. The other half is contributed by the 
increase of $\lambda^{-2}$, which controls the extent to which off diagonal elements of $\hat\rho$ are negligible. Therefore, in the dilute regime (but at high enough energy where $e^{\gamma y}\gg 1$) the entropy grows due to two distinct effects: growth of the dimension $d$ of the subspace on which $\hat\rho$ is nonvanishing, as well as further decoherence of $\hat\rho$ on this subspace. The two effects contribute equally to the entropy. 

 We note that we did obtain $S_e\approx \frac{\gamma}{2}y$ in the LT  regime, and one might think that in this saturated regime the previous argument holds. However, a closer inspection shows that this is not the case. In the saturation regime, just like before the dimension of the relevant Hilbert space on which the density matrix is close to unity is controlled by the parameter $\mu^2$. However $\mu^2$ now grows with energy much slower than exponentially, i.e. Eq.~(\ref{mu}). Thus, the ``expansion" of the populated subspace of the Hilbert space does not contribute any linear in rapidity term to the entropy. On the other hand, the growth of $\lambda^{-2}$ is still exponential just like in the BFKL regime, Eq.~(\ref{lambda}). All the entropy evolution in Eq.~(\ref{enev}) therefore originates from further decoherence on the Hilbert space of approximately fixed dimension. Therefore, neither in the dilute nor in the dense regime, the linear growth of entropy discussed in the present paper seems to originate entirely from a picture proposed in Ref.~\cite{Kharzeev:2017qzs}, although this conclusion could be basis dependent.


As the last point in this paper, we have also defined the Wigner functional associated with the density matrix $\hat\rho$. We have argued that it can give access to understanding the joint probability distribution in the space of color charge density and color current density.  We hope that such a distribution  can teach us about the momentum distribution of produced gluons in events with high multiplicity, which should be instrumental in understanding the correlated behavior of produced hadrons. We have shown that even within the simple Gaussian ansatz, the distributions of color charges and color currents do not factorize, and that this non factorization feature is not eliminated by high energy evolution.

We hope that further work on these subjects will lead to a better, and more complete  understanding of hadronic physics at high energies.

\acknowledgements
NA thanks Carlos Pajares for discussions on this subject. NA and FD were
was
supported by Ministerio de Ciencia e
Innovaci\'on of Spain under project FPA2017-83814-P and Unidad de Excelencia Mar\'{\i}a de
Maetzu under project MDM-2016-0692, by Xunta de Galicia (Conseller\'{\i}a de Educaci\'on) within the Strategic Unit
AGRUP2015/11, and by FEDER. AK was supported by the NSF Nuclear Theory grant 1614640. ML was supported by the Israeli Science Foundation grant \#1635/16; AK and ML were also supported by the BSF grant \#2014707. This work has been performed
in the framework of COST Action CA15213 ``Theory of hot matter and relativistic heavy-ion collisions" (THOR).

\bibliography{bibliography}

\end{document}